\documentclass[a4paper,11pt]{article}
\pdfoutput=1 

\usepackage{jheppub} 
\usepackage[T1]{fontenc} 
\usepackage{float}
\usepackage[usenames,dvipsnames]{xcolor}
\usepackage{mathtools}
\usepackage{graphicx}
\usepackage{subcaption} 
\usepackage{tikz}
\usetikzlibrary{fit,positioning}

\usepackage{colortbl}
\usepackage{xcolor}
\definecolor{LightCyan}{rgb}{0.88,1,1}

\usepackage{multirow}
\usepackage{booktabs}

\usepackage{multirow}
\usepackage{caption} 

\usepackage{amsfonts,epsf,simplewick, epsf,feynmp}
\usepackage{booktabs}
\usepackage{cancel}
\usepackage{braket}
\usepackage{array}
\usepackage{rotating}
\usepackage{bm}
\usepackage{arydshln}
\usepackage{epsfig}
\usepackage{epstopdf}

\usepackage{tikz}
\usepackage{quantikz}
\usepackage{tikz-3dplot}
\usetikzlibrary{decorations.pathreplacing}
\usepackage{indentfirst}
\def\uppertr{\!\!\!\!\scalebox{0.6}{\mbox{
\setlength{\unitlength}{2pt}
\begin{picture}(5,5)
\put(0,5){\thicklines\line(1,0){5}}  
\put(5,0){\thicklines\line(0,1){5}}  
\put(0,5){\thicklines\line(1,-1){5}}  
\end{picture}}}}
\def\suppertr{\!\!\!\scalebox{0.4}{\mbox{
\setlength{\unitlength}{2pt}
\begin{picture}(5,5)
\put(0,5){\thicklines\line(1,0){5}}  
\put(5,0){\thicklines\line(0,1){5}}  
\put(0,5){\thicklines\line(1,-1){5}}  
\end{picture}}}}

\title{Fun with Graph States:\\ Nonlocal Bell pairs and the Arf Invariant}

\author{Bartłomiej Czech,}
\author{Yichen Feng,}
\author{Xianlai Wu}
\author{and Minjun Xie}

\affiliation{Institute for Advanced Study, Tsinghua University, Beijing 100084, China}

\emailAdd{bartlomiej.czech@gmail.com}
\emailAdd{fyc23@mails.tsinghua.edu.cn}
\emailAdd{wuxl24@mails.tsinghua.edu.cn}
\emailAdd{xiemj21@mails.tsinghua.edu.cn}

\abstract{
We study inner products and partial amplitudes of graph states---a commonly employed class of quantum states, which are specified by graphs. We find that the magnitudes of these quantities are simply related to the rank of the adjacency matrix of the graph over $\mathbb{F}_2$ while the phase is determined by the Arf invariant of its quadratic refinement. These facts motivate a nonlocal tensor factorization of the Hilbert space, with respect to which all graph states are products of Bell pairs with unentangled ancillae. These results may illuminate the quantum advantage in the framework of Measurement-Based Quantum Computation and suggest that graph states can be usefully visualized in the language of algebraic topology. In addition, we develop a specialized technique for computing expectation values of qubit-wise permutations in graph states, which is useful for calculating multi-invariants.
}

\begin{document} 
\maketitle
\flushbottom

\section{Introduction}
Stabilizer states and, more specifically, graph states (reviewed in Section~\ref{sec:review}) are among the most thoroughly studied quantum states. Originally defined for the purpose of constructing quantum error correcting codes \cite{Schlingemann_2001, schlingemann2001stabilizercodesrealizedgraph}, they were later adopted as resource states for Measurement-Based Quantum Computation (MBQC) \cite{PhysRevLett.86.5188, PhysRevA.68.022312}. Beyond these two core uses, their highly ordered structure has made them useful in a broad variety of contexts---from exhibiting the strongest Bell-type inequalities \cite{Baccari2018Bell,Zhao:2020fsy} to studying black hole evaporation \cite{Hayden:2007cs,Yoshida:2021xyb}. 
This paper is concerned with an aspect of graph states, which is relevant to all the above applications: the inner product $\braket{G'|G''}$ of two graph states.

As is easy to see---and as we explain in Section~\ref{sec:innerproduct}---the inner product $\braket{G'|G''}$ reduces to the partial amplitude $\braket{+|^{\otimes M} | G}$, where $\ket{+}$ is the $(+1)$-eigenstate of the Pauli $X$ operator and $G$ is another graph, which is simply related to graphs $G'$ and $G''$. Consequently, our paper focuses on the partial amplitude $\braket{+|^{\otimes M} | G}$, though in Appendix~\ref{app:wfx} we include the simple generalization to partial amplitudes of $\ket{G}$ in the full Pauli-$X$ eigenbasis. To our mild surprise, dedicated studies of quantity $\braket{+|^{\otimes M} | G}$ are nearly absent in the literature. Two exceptions are papers \cite{math11102310} and \cite{akella2026multiinvariantsstabilizerstates}, on which we comment below.

What is even more surprising is that quantity $\braket{+|^{\otimes M} | G}$ harbors a wealth of mathematical connections and physical insights, which our paper only begins to explore:
\begin{enumerate}
\item The magnitude of $\braket{+|^{\otimes M} | G}$ is determined by the rank (over the binary field $\mathbb{F}_2$) of the adjacency matrix of graph $G$. As we sketch in the Discussion, this fact is related to the computational power of $\ket{G}$ when it is used as a resource in Measurement-Based Quantum Computation.
\item The phase of $\braket{+|^{\otimes M} | G}$ is determined by the Arf invariant \cite{Arf1941,Dye1978} of a quadratic refinement of the adjacency matrix of graph $G$. In geometry, the Arf invariant characterizes spin structures on two-dimensional manifolds; in physics, it counts (mod 2) zero modes of the Dirac operator. In the Discussion, we comment on the likely significance of the appearance of the Arf invariant in these diverse contexts.
\item Facts~1. and 2. motivate a new basis of \emph{effective} qubits, which reorganizes the physical Hilbert space where $\ket{G}$ lives. In the effective basis, \emph{every} graph state $\ket{G}$ is a product of unentangled qubit states $\ket{\pm}$ and Bell pairs; see Figure~\ref{appfig:new_graph}. We emphasize that the effective qubits form an alternative tensor product decomposition of the physical Hilbert space. The relation of the effective qubits to the physical qubit constituents of $\ket{G}$ is, for generic graphs $G$, highly nonlocal. 
\end{enumerate}
After the research for this paper was completed, we discovered that results, which are mathematically equivalent to points~1. and 2. had previously appeared in \cite{math11102310}, albeit in a very different form. We hope that readers who are already familiar with \cite{math11102310} will benefit from the physical insight, which is contained in and implied by point 3.

In our view, results~1.-3. offer a gateway to a richer understanding of graph states and, possibly, of quantum computing. The appearance of the Arf invariant suggests that graph states can be usefully visualized using the language of algebraic topology. We sketch a preliminary picture in the Discussion but defer a complete presentation to a future publication. In the present paper, we concentrate on a more immediate, physical message: that a nonlocal change of basis (an unconventional tensor decomposition of the Hilbert space) brings all graph states to the same, nearly trivial form shown in Figure~\ref{appfig:new_graph}. Graph states being resources for Measurement-Based Quantum Computation, this insight suggests a novel perspective on the issue of the quantum advantage: what exactly makes quantum computing more powerful than its classical forerunner? Indeed, if measuring the ``simple'' states shown in Figure~\ref{appfig:new_graph} can simulate any quantum computation, whence the computational enhancement? We propose a likely answer in the Discussion, leaving a full exploration of this question to a follow-up. 

A recent paper \cite{akella2026multiinvariantsstabilizerstates}\footnote{We thank the authors of \cite{akella2026multiinvariantsstabilizerstates} for sharing a manuscript of their paper prior to its arXiv announcement.} studies the inner product of two graph states toward an entirely different end: to probe a family of quantities known as multi-invariants, which generalize R{\'e}nyi entropies and which have recently attracted attention in the context of the AdS/CFT correspondence \cite{Gadde_2022, Penington_2023, Gadde2025}. In an effort to meet all demands for calculating $\braket{G'|G''}$, we devote Section~\ref{sec:compute_permut} to this specific application, which imposes an additional structure on graphs $G'$ and $G''$. While fact~2.~extends the results of \cite{akella2026multiinvariantsstabilizerstates}, we find that for practical purposes an altogether different approach based on simple group theory is often more handy. This alternative approach---and an explanation of how it complements and reproduces our main line of calculations---comprises Section~\ref{sec:compute_permut}. That section is logically independent from the rest of the paper and a reader unconcerned with multi-invariants can skip it at no peril. 

\paragraph{Organization} To make the paper self-contained, we briefly review stabilizer states and graph states in Section~\ref{sec:review}. In Section~\ref{sec: Innerprod} we calculate the magnitude of $\bra{+}^{\otimes M}\ket{G}$ and offer its operational interpretation. We then calculate the phase of $\bra{+}^{\otimes M}\ket{G}$ in Section~\ref{sec: Arf_inv} and find that it is simply related to the Arf invariant. Mathematically, the calculation in Section~\ref{sec: Arf_inv} subsumes the results of Section~\ref{sec: Innerprod}, but its physical significance is quite different. In particular, it motivates the nonlocal basis of effective qubits, which we view as the main value added of this paper. Calculations pertaining to multi-invariants comprise Section~\ref{sec:compute_permut}. In Section~\ref{sec:discuss} we sketch a likely quantum-computational and geometric interpretation of our results. The appendices contain a computation of amplitudes such as $\braket{+--\ldots | G}$ and an assortment of multi-invariant calculations.

\paragraph{Notation and conventions}
We represent the action of the Pauli group on a single qubit Hilbert space with the following matrices:
\begin{equation}
    I:=\begin{bmatrix}
        1 & 0 \\
        0 & 1
    \end{bmatrix}\quad 
    X:=\begin{bmatrix}
        0 & 1 \\
        1 & 0
    \end{bmatrix}\quad 
    Y:=\begin{bmatrix}
        0 & -i \\
        i & \phantom{-}0
    \end{bmatrix}\quad 
    Z:=\begin{bmatrix}
        1 & \phantom{-}0 \\
        0 & -1
    \end{bmatrix}.
\end{equation}
On many-qubit Hilbert spaces, we carefully distinguish matrix multiplication from tensor products with the absence (respectively, presence) of the sign $\otimes$. Where necessary, the qubit on which an operator acts is indicated with a subscript. For example, $X_1\otimes Z_2$ acts on a two-qubit Hilbert space while $XZ$ is an operator on a single qubit Hilbert space. 

We reserve $I$ for the identity operator acting on a single qubit Hilbert space while $\mathbb{I}$ is the identity on a generic many-qubit Hilbert space. We also operate on $\mathbb{F}_2^M$ as a vector space, where $\mathbb{F}_2$ is the Galois field with two elements. The identity operator acting on $\mathbb{F}_2^M$ is denoted with 1.

\section{Brief review of graph states} 
\label{sec:review}
Graph states are a particularly convenient subclass of the broader category of {\bf stabilizer states}. A good review of stabilizer states is found in \cite{Nielsen_Chuang_2010}.

An $M$-qubit stabilizer state $\ket{\Psi}$ is the unique common eigenstate of $M$ mutually commuting operators $K_i$ ($i=1,\dots,M$) in the Pauli group. In defining stabilizer states, most references set all the eigenvalues to $+1$.  The $M$ commuting operators are called the {\bf stabilizers} of $\ket{\Psi}$. Consider the projections
\begin{equation}
	P_i=\frac{\mathbb{I}+K_i}{2}, \label{eqn:proj}
\end{equation}
which map into the (+1)-eigenspaces of $K_i$. It is easy to see that the quantum state $\ket{\Psi}$ satisfies 
\begin{equation} 
\label{eq:rho_stab}
    \ket{\Psi}\bra{\Psi}=\prod_{i=1}^M P_i=\frac{1}{2^M}\sum_{g\,\in\, \mathcal{G}}\,g,
\end{equation}
where $\mathcal{G}$ is the {\bf stabilizer group} generated by $K_1,K_2,\dots K_M$.

\paragraph{Graph states}
A major virtue of graph states is that every stabilizer state is locally Clifford (LC)-equivalent to at least one graph state \cite{PhysRevA.69.022316}. A comprehensive review of graph states is \cite{Hein2006}.

An $M$-qubit graph state $\ket{G}$ is associated to a simple undirected graph $G$ whose vertices correspond to constituent qubits. (A graph is simple if it has no loops and at most one edge between any two vertices.) Denoting the $X$-eigenbasis with $X\ket{\pm} = \pm \ket{\pm}$, the graph state $\ket{G}$ is:
\begin{equation}
\ket{G}\equiv\prod_{i \sim j}CZ_{i,j}\ket{+}^{\otimes M}. 
\label{eq:def_graph} 
\end{equation}
Here $i \sim j$ signifies that an edge connects the underlying vertices while $CZ_{i,j}$ is the {\bf controlled-$Z$} gate, which acts on the $Z$-eigenbasis of the two-qubit Hilbert space as:
\begin{equation}
\label{defcz}
    \ket{00} \to \ket{00} \qquad 
    \ket{01} \to \ket{01} \qquad 
    \ket{10} \to \ket{10} \qquad 
    \ket{11} \to -\ket{11} .
\end{equation}
In short, the graph represents the pattern of $CZ$ gates, which prepare $\ket{G}$ from $\ket{+}^{\otimes M}$.

It follows from 
\begin{equation}
\label{czconjugation}
    CZ_{i,j}\, \big(I_i \otimes X_j\big)\, CZ_{i,j} = Z_i \otimes X_j 
\end{equation}
that $\ket{G}$ is a stabilizer state whose stabilizer group is generated by:
\begin{equation}
K_j \equiv \left(\bigotimes_{i \sim j} Z_i\right) \otimes X_j
\qquad {\rm for}~1 \leq j \leq M .
\label{graphstabgen}
\end{equation}
Equations~(\ref{eq:def_graph}) and (\ref{graphstabgen}) are nicely summarized using the adjacency matrix of graph $G$ as well as its upper-diagonal part:
    \begin{equation}
    \label{defa}
        A_{ij}\,=\,\begin{cases}
            1\quad {\rm if}~i \sim j\\
            0\quad \text{otherwise}
        \end{cases}
        \qquad \quad
        A^{\uppertr}_{ij}\,=\,\begin{cases}
            1\quad {\rm if}~i \sim j~~{\rm and}~~i < j\\
            0\quad \text{otherwise}
        \end{cases} .
    \end{equation}
In terms of $A$ and $A^{\uppertr}$, we have:
\begin{align}
\label{GintermsofA}
\ket{G} & = 
\prod_{i,j=1}^M \left(CZ_{i,j}\right)^{A^{\suppertr}_{ij}}\ket{+}^{\otimes M} ,
\\
\label{KintermsofA}
K_j & 
= \left(\bigotimes_{i \sim j} \left(Z_i\right)^{A_{ij}}\right)\otimes X_j .
\end{align}
Note that (\ref{GintermsofA}) uses the upper diagonal matrix $A^{\uppertr}$ because an edge $i \sim j$ indicates one application of $CZ_{i,j}$, not two. The distinction between $A$ and $A^{\uppertr}$ will gain an uncanny importance in Section~\ref{sec: Arf_inv}. 

Because $(CZ_{i,j})^2=I_i\otimes I_j$, it is natural to view (\ref{defa}) as belonging to $\mathbb{F}_2$, the Galois field with two elements. Therefore, throughout this paper, we take $A$ and $A^{\uppertr}$ and the vector spaces they act upon to be defined over $\mathbb{F}_2$. 

\subsection{Inner product of two graph states as an amplitude}
\label{sec:innerproduct}
We are interested in the inner product $\braket{G'|G''}$, where both $\ket{G'}$ and $\ket{G''}$ are $M$-qubit graph states. In graph-theoretic language, $G'=(V,E')$ and $G''=(V,E'')$ are two simple and undirected graphs, which share the same vertex set $V$ but have distinct edge sets $E'$ and $E''$. The inner product $\braket{G'|G''}$ is then computed by an amplitude of another graph $G=(V,E)$:
\begin{equation}
    \braket{G'|G''}=\bra{+}^{\otimes M} \underbrace{\left(
        \prod_{(i',j')\in E'} CZ_{i',j'}
    \right)\left(
        \prod_{(i'',j'')\in E''} CZ_{i'',j''}
    \right)
    \ket{+}^{\otimes M}}_{\ket{G}}
    \equiv
    \bra{+}^{\otimes M}\ket{G}.
\end{equation}  
Let $A'$ and $A''$ be the adjacency matrices of $G'$ and $G''$. Because $(CZ_{i,j})^2=I_i\otimes I_j$, the adjacency matrix $A$ of $G$ is $A=A'+A''$ (over $\mathbb{F}_2$). From here on, we focus on quantity $\bra{+}^{\otimes M}\ket{G}$ for general graphs $G$.

\section{Rank of adjacency matrix determines the magnitude}
\label{sec: Innerprod}
 
\subsection{Overview of results}
\label{graphcalcoverview}
This section focuses exclusively on the magnitude of $\bra{+}^{\otimes M}\!\ket{G}$; the $\pm$ phase is computed in Section~\ref{sec: Arf_inv}. We derive the following:
\begin{enumerate}
  \item $\bra{+}^{\otimes M}\ket{G}=0$ if and only if $G$ contains an  \textit{induced subgraph}\footnote{An induced subgraph $H$ of $G$ is a graph, which comprises a subset of the vertices of $G$ and \emph{all} $G$-edges that connect them  \cite{diestel2005graph}.}
  $H$ such that every vertex in $G$ has an even number of neighbors in $H$ and $H$ contains an odd number of edges.
  \item If $\bra{+}^{\otimes M}\ket{G}\neq 0$ then
  \begin{equation}
  \label{eq: rank G formula}
        \left|\bra{+}^{\otimes M}\ket{G}\right|^2
        =2^{-\text{\rm rank} A} .
    \end{equation}
\end{enumerate}

The calculations are organized as follows. In Subsection~\ref{sec:therankstory} we prove Result~2. In Subsection~\ref{sec:homomorphism} we recognize Result~1 as a corollary of the proof of Result~2. Subsection~\ref{sec:operational} explains the results in operational terms. 

\subsection{Calculation of the magnitude}
\label{sec:therankstory}
We wish to prove equation~\eqref{eq: rank G formula}.
We remind the reader that matrix $A$ and all associated concepts are defined over $\mathbb{F}_2$ and that is how we should understand the rank in (\ref{eq: rank G formula}). Using equation~(\ref{eq:rho_stab}), we must compute:
\begin{equation}
\label{eq: absolute square of inner product}
    \left|\bra{+}^{\otimes M}\ket{G}\right|^2
    =\frac{1}{2^{M}}\sum_{g\in \mathcal{G}} 
    \bra{+}^{\otimes M} g \ket{+}^{\otimes M} .
\end{equation}
Each stabilizer $g$ is a tensor product of $I$, $X$, $Z$ and $XZ$, potentially dressed with a phase. Since $\bra{+} Z \ket{+} = \bra{+} XZ \ket{+} = 0$, however, nonvanishing summands in (\ref{eq: absolute square of inner product}) come from those $g \in \mathcal{G}$, which only involve $I$ and $X$. As we presently explain, such stabilizers $g$ are in one-to-one correspondence with:
\begin{itemize}
    \item vectors in $\ker A \subset \mathbb{F}_2^M$;
    \item induced subgraphs $H$ of $G$ such that every vertex in $G$ has an even number of neighbors in $H$.
\end{itemize}

In equation~(\ref{KintermsofA}) we saw that the columns of $A$ correspond to generators of $\mathcal{G}$. In particular, the components of the $j^{\rm th}$ column indicate the locations of $Z_i$ factors in $K_j$. By the same token, the matrix product $A \vec{x}$ (where $\vec{x} \in \mathbb{F}_2^M$) represents the locations of $Z$ or $XZ$ factors in the stabilizer $O_{\vec{x}} = \prod_{i | x_i = 1} K_i$, which is explicitly given by:
\begin{equation}
    O_{\vec{x}} =
    \pm X_1^{x_1} Z_1^{(A\vec{x})_1} \otimes X_2^{x_2} Z_2^{(A\vec{x})_2} \otimes \ldots \otimes X_M^{x_M} Z_M^{(A\vec{x})_M}.
    \label{arbitrarystab}
\end{equation}
A possible sign arises from applications of $ZXZ = -X$ and we study it in Subsection~\ref{sec:homomorphism}. For the present purposes, we observe that stabilizer (\ref{arbitrarystab}) features no $Z$ or $XZ$ factors if and only if $A\vec{x}=0$. This confirms the first characterization of nonvanishing terms in (\ref{eq: absolute square of inner product}).

Solutions of $A\vec{x}=0$ are also bijectively related to the special induced subgraphs of $G$, which we described above. That is because each vector $\vec{x} \in \mathbb{F}_2^M$ can be interpreted as the indicator function of an induced subgraph $H_{\vec{x}}$:
\begin{equation}
\label{graphindicator}
    x_i = \begin{cases}
            1\quad \textrm{if the $i^{\rm th}$ vertex of $G$ is in $H_{\vec{x}}$}\\
            0\quad \text{otherwise}
        \end{cases}.
\end{equation}
In that reading, $(A\vec{x})_j$ tells us whether the $j^{\rm th}$ vertex of $G$ has an even or an odd number of neighbors in $H_{\vec{x}}$. We have $A\vec{x} = 0$ if and only if every vertex in $G$ has an even number of neighbors in $H_{\vec{x}}$.

In summary, we have thus far found
\begin{equation}
\label{eq: analytic form of the square of the inner product}
  \left|\bra{+}^{\otimes M}\ket{G}\right|^2
  =
  \frac{1}{2^M}
  \sum_{\vec{x}\,\in\, \ker A}(-)^{\phi(\vec{x})},
\end{equation}
where $\phi(\vec{x}) \in \mathbb{F}_2$ is \emph{defined} by equation\footnote{We give an explicit formula for $\phi(\vec{x})$ in equation~\eqref{phiexplicit1}. That formula extends $\phi$ to arbitrary $\vec{x}\in\mathbb{F}_2^M$, but in this subsection we focus on $\vec{x}\in \ker A$.} 
\begin{equation}
    O_{\vec{x}} 
    \equiv (-)^{\phi(\vec{x})}\bigotimes_{i | x_i = 1} X_i.
    \label{kernelstab}
\end{equation}
Equation~(\ref{kernelstab}) is (\ref{arbitrarystab}) with $A\vec{x}=0$. 

Consider $\ker A$ as a group under addition over $\mathbb{F}_2$. Then $\phi(\vec{x})$ is a group homomorphism from $\ker A$ to $\mathbb{Z}_2$ because:
\begin{align}
(-)^{\phi(\vec{x}+\vec{y})} \bigotimes_{k | (x+y)_k = 1} X_k \equiv O_{\vec{x}+\vec{y}} = O_{\vec{x}} O_{\vec{y}} 
& = (-)^{\phi(\vec{x})}\bigotimes_{i | x_i = 1} X_i
\, \cdot \,
  (-)^{\phi(\vec{y})}\bigotimes_{j | y_j = 1} X_j 
  \\
& =
(-)^{\phi(\vec{x})+\phi(\vec{y})} \bigotimes_{k | (x+y)_k = 1} X_k ,
\end{align}
so
\begin{equation}
    \phi(\vec{x}+\vec{y}) = \phi(\vec{x}) + \phi(\vec{y}).
    \label{confirmhomom}
\end{equation}
Now the first isomorphism theorem tells us that $\phi: \ker A \to \mathbb{Z}_2$ can only be structured in one of two ways:
\begin{itemize}
    \item[(A)] Exactly half the elements of $\ker A$ are mapped by $\phi$ to 0 and the other half to 1.
    \item[(B)] $\phi(\vec{x}) = 0 \quad \forall\, \vec{x} \in \ker A$.
\end{itemize}
In Case~(A), equation~(\ref{eq: analytic form of the square of the inner product}) evaluates to 0. In Case~(B), equation~(\ref{eq: analytic form of the square of the inner product}) boils down to:
\begin{equation}
     \left|\bra{+}^{\otimes M}\ket{G}\right|^2
  = 2^{-M} \cdot 2^{\dim \ker A} = 2^{-{\rm rank} A}.
  \label{rankformularepeated}
\end{equation}
This proves equation~(\ref{eq: rank G formula}) and Result~2 highlighted in Subsection~\ref{graphcalcoverview}. 

\subsection{Quantity $\phi(\vec{x})$ counts edges in induced subgraphs}
\label{sec:homomorphism}

Returning to Case~(A), we see that $\bra{+}^{\otimes M}\ket{G}=0$ if and only if some stabilizer of $\ket{G}$ is built entirely of $X$-operators and carries a $(-1)$ phase. Comparing with equation~(\ref{kernelstab}), this says that there exists at least one $\vec{x} \in \ker A$ with $\phi(\vec{x})=1$. We presently explain that $\phi(\vec{x})$ counts (mod 2) edges internal to the induced subgraph $H_{\vec{x}}$, for which $\vec{x}$ is the indicator function; see equation~(\ref{graphindicator}). 

Equation~(\ref{GintermsofA}) implies that (\ref{kernelstab}) is:
\begin{equation}
\mathcal{O}_{\vec{x}} = 
\prod_{i,j=1}^M \left(CZ_{i,j}\right)^{A^{\suppertr}_{ij}}
\left(\bigotimes_{k | x_k = 1} X_k \right)
\prod_{i,j=1}^M \left(CZ_{i,j}\right)^{A^{\suppertr}_{ij}},
\label{oxinterm}
\end{equation}
where $A^{\uppertr}$ is the upper diagonal part of the adjacency matrix; see~(\ref{defa}). We noted in (\ref{czconjugation}) that conjugation by $CZ$ gates dresses a product of $X$s with factors of $Z$. The $CZ_{i,j}$ gates for which $x_i=x_j=0$ drop out from (\ref{oxinterm}) because they commute past $\bigotimes_k X_k^{x_k}$. Those for which $x_i \neq x_j$ drop out when $A\vec{x}=0$, as can be seen from equation~(\ref{arbitrarystab}). We therefore find that operator~(\ref{oxinterm}) can be rewritten as:
\begin{equation}
\mathcal{O}_{\vec{x}} = 
\prod_{i|x_i=1} \prod_{j|x_j=1} \left(CZ_{i,j}\right)^{A^{\suppertr}_{ij}}
\left(\bigotimes_{k | x_k = 1} X_k \right)
\prod_{i|x_i=1} \prod_{j|x_j=1} \left(CZ_{i,j}\right)^{A^{\suppertr}_{ij}}
\label{oxinterm2}.
\end{equation}
From~(\ref{czconjugation}), conjugation by $CZ_{i,j}$ gates has the following effect:
\begin{align}
    CZ_{i,j}\, \big( X_i \otimes X_j\big)\, CZ_{i,j} 
    & \quad\longleftrightarrow\quad 
    - X_i Z_i \otimes X_j Z_j \nonumber, \\
    CZ_{i,j}\, \big( X_i Z_j \otimes X_j\big)\, CZ_{i,j} 
    & \quad\longleftrightarrow\quad 
    - X_i \otimes X_j Z_j.
    \label{czonxx}
\end{align}
Again, if $A\vec{x}=0$ then the final form of $\mathcal{O}_{\vec{x}}$ is $(-1)^{\phi(\vec{x})} \otimes_k X^{x_k}$ so we need not keep track of individual factors of $Z$, only the overall sign in front. But equations~(\ref{czonxx}) tell us that \emph{every} conjugating $CZ_{i,j}$ gate in (\ref{oxinterm2}) brings a single $(-1)$ factor to $\phi(x)$. This implies: 
\begin{equation}
    \phi(\vec{x}) 
    = \sum_{i | x_i=1} \sum_{j | x_j=1} A^{\uppertr}_{ij}
    = \vec{x}^\intercal A^{\uppertr} \vec{x}.
\label{phiexplicit1}
\end{equation}
Although we have focused on $\vec{x}\in \ker A$, there is no obstruction to extending the definition of $\phi$ in this way to arbitrary $\vec{x}\in \mathbb{F}_2^M$. Doing so will be useful in Section~\ref{sec: Arf_inv}.

From the definition of an induced subgraph $H_{\vec{x}}$ in (\ref{graphindicator}), equation~(\ref{phiexplicit1}) is identical to:
\begin{equation}
    \phi(\vec{x}) = 
    \sum_{i < j} 
    \delta\big(\textrm{vertices $i$ and $j$ both belong to $H_{\vec{x}}$}\big)\,
    \delta\big(i \sim j\big)
\label{phiexplicit2}.
\end{equation}
That is, $\phi(\vec{x})$ tells us whether $H_{\vec{x}}$ contains an even or an odd number of edges, as claimed. 

\subsection{An operational interpretation}
\label{sec:operational}

We wish to give an operational interpretation of equation~(\ref{rankformularepeated}). Quantity~$|\bra{+}^{\otimes M}\ket{G}|^2$ is the probability of measuring the graph state $\ket{G}$ in the $X$-eigenbasis and finding $+1$ on every qubit. We now explain on physical grounds that this probability must be either 0 or (\ref{rankformularepeated}).

\paragraph{A convenient set of generators of the stabilizer group}
We commented that columns of $A$ define a set of generators $K_i$ for the stabilizer group of $\ket{G}$, as captured by equation~(\ref{KintermsofA}). Formula~(\ref{arbitrarystab}) allows us to exhibit other, more convenient sets of generators by transforming $A$ with column operations familiar from linear algebra. Let us use (a slightly unorthodox version of) Gaussian elimination to bring $A$ to `bottom-aligned column echelon form' $C$. As matrix multiplication, this operation can be expressed as $A\, V = C$, or in column components:  
\begin{equation}
\label{gaussianschematic}
    A\,
    \left[\begin{array}{cccccc}
    \vdots & \vdots & \vdots & \vdots & & \vdots\\
    \phantom{0} & \phantom{0} & \phantom{0} & \phantom{0} & \phantom{0} & \phantom{0}\\
    \vec{v}_1 & \vec{v}_2 & \vec{v}_3 & \vec{v}_4 & ~\cdots~ & \vec{v}_M\\
    ~\phantom{0}~ & ~\phantom{0}~ & ~\phantom{0}~ & ~\phantom{0}~ & ~\phantom{0}~ & ~\phantom{0}~ \\
    \vdots & \vdots & \vdots & \vdots & & \vdots
        \end{array}\right]
=
    \left[\begin{array}{cccccc}
    ~{\color{red} 1}~ & ~0~ & ~0~ & ~0~ & ~\cdots~ & ~0~\\
    {\rm x} & 0 & 0 & 0 & \cdots & ~0~\\
    {\rm x} & 0 & 0 & 0 & \cdots & ~0~\\
    {\rm x} & ~{\color{red} 1}~ & 0 & 0 & \cdots & ~0~\\
    {\rm x} & {\rm x} & 0 & 0 & \cdots & ~0~\\
    {\rm x} & {\rm x} & ~{\color{red} 1}~ & 0 & \cdots & ~0~
    \end{array}\right]
\end{equation}
In~(\ref{gaussianschematic}), we chose $C$ to illustrate the `bottom-aligned column echelon' property; the entries marked `x' are arbitrary. As a second step, we reorder the qubits so that the topmost entry in each non-zero column of $C$ lands on the diagonal. This is done by left-multiplying equation~(\ref{gaussianschematic}) with some permutation matrix $R$. Redefining $A \to R A R^{-1}$, $V \to RV$, and  $C \to RC$, we obtain a new equation ${A}\,{V} = {C}$, now in the form: 
\begin{equation}
\label{gaussianschematic2}
    A\,
    \left[\begin{array}{cccccc}
    \vdots & \vdots & \vdots & \vdots & & \vdots\\
    \phantom{0} & \phantom{0} & \phantom{0} & \phantom{0} & \phantom{0} & \phantom{0}\\
    \vec{v}_1 & \vec{v}_2 & \vec{v}_3 & \vec{v}_4 & ~\cdots~ & \vec{v}_M\\
    ~\phantom{0}~ & ~\phantom{0}~ & ~\phantom{0}~ & ~\phantom{0}~ & ~\phantom{0}~ & ~\phantom{0}~ \\
    \vdots & \vdots & \vdots & \vdots & & \vdots
        \end{array}\right]
=
    \left[\begin{array}{cccccc}
    ~{\color{red} 1}~ & ~0~ & ~0~ & ~0~ & ~\cdots~ & ~0~\\
    {\rm x} & ~{\color{red} 1}~ & 0 & 0 & \cdots & ~0~\\
    {\rm x} & {\rm x} & ~{\color{red} 1}~ & 0 & \cdots & ~0~\\
    {\rm x} & {\rm x} & {\rm x} & 0 & \cdots & ~0~\\
    {\rm x} & {\rm x} & {\rm x} & 0 & \cdots & ~0~\\
    {\rm x} & {\rm x} & {\rm x} & 0 & \cdots & ~0~
    \end{array}\right]
\end{equation}
We stress that transformation $A \to R A R^{-1}$ does not affect the state $\ket{G}$ except by relabeling its constituent qubits. In the remainder of this section we assume that the qubits are labeled $1 \ldots M$ in the way dictated by (\ref{gaussianschematic2}). We further remark that the number of nonvanishing columns in $C$ equals the rank of $A$ because $C$ was obtained from a `column echelon form' of $A$.

According to formula~(\ref{arbitrarystab}), the columns of $V$ and $C$ in (\ref{gaussianschematic2}) determine a new set of generators $O_{\vec{v}_1},\, O_{\vec{v}_2}, \ldots O_{\vec{v}_M}$. This set of generators enjoys the following key properties, which follow directly from formula~(\ref{arbitrarystab}):
\begin{itemize}
    \item $C$ being lower triangular implies:
    \begin{equation}
    \label{gen1}
    [X_k, O_{\vec{v}_{k'}}]=0 \quad \forall\, k' > k
\end{equation}
        \item The 1's on the diagonal imply:
    \begin{equation}
    \label{gen2}
    \{X_k, O_{\vec{v}_k}\}=0 
    \quad {\rm for}~~1 \,\leq\, k \,\leq\, {\rm rank} A \phantom{+ 1}
\end{equation}
        \item The 0's on the diagonal imply:
    \begin{equation}
    \label{gen3}
\,\,\,\,\,\,\,\,    [X_k, O_{\vec{v}_k}]=0 
    \quad {\rm for}~~{\rm rank} A + 1 \,\leq\, k \,\leq\, M
\end{equation}
\end{itemize}

\paragraph{Qubit-wise measurements}
We now measure the qubits of $\ket{G}$ in the $X$-eigenbasis sequentially, one at a time, in the order assumed in equations~\eqref{gen1}-\eqref{gen3}. We are interested in the probability of measuring $+1$ at each step.

Let $\ket{G} \equiv \ket{\Psi_0}$. If we measure $X_1$ and obtain $+1$, the state becomes $\ket{+} \otimes \ket{\Psi_1}$ for some $(M-1)$-qubit state $\ket{\Psi_1}$. We claim that $\ket{\Psi_1}$ is a stabilizer state whose stabilizing operators are given by formula~(\ref{arbitrarystab}), with the topmost row and leftmost column of $V$ and $C$ erased. Explicitly, $\ket{\Psi_1}$ is stabilized by the following $M-1$ commuting Pauli operators indexed by $2 \leq k \leq M$:
\begin{equation}
    O^{(1)}_{\vec{v}_{k}} =
    \pm
    X_{2}^{(\vec{v}_{k})_{2}} Z_{2}^{(A\vec{v}_{k})_{2}} 
    \otimes 
    X_{3}^{(\vec{v}_{k})_{3}} Z_{3}^{(A\vec{v}_{k})_{3}} 
    \otimes \ldots \otimes
    X_{M}^{(\vec{v}_{k})_{M}} Z_{M}^{(A\vec{v}_{k})_{M}}
\end{equation}
This is true because equation~(\ref{gen1}) guarantees that $\ket{+} \otimes \ket{\Psi_1}$ is an eigenstate of $O_{\vec{v}_{k}}$ and, by construction, we have either $O_{\vec{v}_{k}} = I_1 \otimes O^{(1)}_{\vec{v}_{k}}$ or $O_{\vec{v}_{k}} = X_1 \otimes O^{(1)}_{\vec{v}_{k}}$. Moreoever, any two operators $O^{(1)}_{\vec{v}_{k}}$ and $O^{(1)}_{\vec{v}_{k'}}$ commute because $O_{\vec{v}_{k}} = \big( I_1~{\rm or}~X_1\big) \otimes O^{(1)}_{\vec{v}_{k}}$ and ${O_{\vec{v}_{k'}} =}\big( I_1~{\rm or}~X_1\big) \otimes O^{(1)}_{\vec{v}_{k'}}$ commute. 

Now consider the full sequence of $X$-measurements, one qubit at a time. Let the state obtained after $l$ consecutive $(+1)$ outcomes be $\ket{+}^{\otimes l} \otimes \ket{\Psi_l}$. The above argument applies to all $\ket{\Psi_l}$, with the recursive step guaranteed by equation~(\ref{gen1}). We therefore find:
\begin{align}
    \label{eq: ProbInterpret}
  |\bra{+}^{\otimes M}\ket{G}|^2
  & =\prod_{l=1}^{M} 
  \,\textrm{Prob\Big(measuring~$X_l = +1$~in~state~$\ket{\Psi_{l-1}}$\Big)} 
  \nonumber \\
  & = \prod_{l=1}^M\, 
  {\rm Tr} \left(\frac{I_l + X_l}{2} \ket{\Psi_{l-1}}\bra{\Psi_{l-1}} \right)
\end{align}

For each factor in (\ref{eq: ProbInterpret}) we have two scenarios, depending on the value of $C_{11}$: 
\begin{itemize}
\item If $C_{11} = 1$ then equation~(\ref{gen2}) implies:
\begin{align}
   & {\rm Tr} \left(\frac{I_l + X_l}{2} \ket{\Psi_{l-1}}\bra{\Psi_{l-1}} \right)
= {\rm Tr} \left(\frac{I_l + X_l}{2} O_{\vec{v}_l} \ket{\Psi_{l-1}}\bra{\Psi_{l-1}} O_{\vec{v}_l} \right) \\
=\, & {\rm Tr} \left(\frac{I_l + O_{\vec{v}_l} X_l O_{\vec{v}_l}}{2} \ket{\Psi_{l-1}}\bra{\Psi_{l-1}} \right)
= {\rm Tr} \left(\frac{I_l - X_l}{2} \ket{\Psi_{l-1}}\bra{\Psi_{l-1}} \right)
= \frac{1}{2} \nonumber.
\end{align}
This scenario arises in product~(\ref{eq: ProbInterpret}) $({\rm rank}A)$ times.
\item If $C_{11} = 0$ then $C$ is the null matrix and all stabilizing operators of $\ket{\Psi_l}$ are (tensor products of) $X$ operators, possibly up to a phase. We previously encountered these stabilizers in equation~(\ref{kernelstab}). We have two sub-scenarios to consider:
\begin{itemize}
    \item If one or more stabilizers of $\ket{\Psi_l}$ carry a $(-1)$ phase then the state is $\ket{+}^{\otimes M-l-q} \otimes \ket{-}^{\otimes q}$ for some $q \geq 1$, or a permutation thereof. Then the probability of measuring $(+1)$ on all its remaining qubits is 0. This is captured by Case~(A) in Subsection~\ref{sec:therankstory} and 
    Result~1 highlighted in Subsection~\ref{graphcalcoverview}. 
    \item If no stabilizers of $\ket{\Psi_l}$ carry a $(-1)$ phase then $\ket{\Psi_l} = \ket{+}^{\otimes M-l}$ and the probability of measuring $(+1)$ on its remaining qubits is 1.
\end{itemize}
\end{itemize}
Substituting these facts in (\ref{eq: ProbInterpret}) recovers Result~2. We now understand the factors of $1/2$ in (\ref{rankformularepeated}) as reflecting probabilities of measuring $X_l = 1$ on individual qubits.

\section{Arf invariant determines the phase} 
\label{sec: Arf_inv}

\subsection{Overview}
\label{Arfcoverview}
If the magnitude of $\bra{+}^{\otimes M}\ket{G}$ is nonzero, it remains to determine the phase. In this section we show that this phase is related to the Arf invariant of the quadratic refinement $\phi(\vec{x})$ of the adjacency matrix $A$, understood as a bilinear form. We explain the relevant concepts as we encounter them.

Recall that gate $CZ$ acts diagonally in the $Z\otimes \ldots \otimes Z$ eigenbasis; viz. equation~(\ref{defcz}). In what follows, it will be convenient to denote $Z\otimes \ldots \otimes Z$ eigenvectors with $\ket{\vec{x}}$ where $\vec{x}\in \mathbb{F}_2^M$. In this notation
\begin{equation}
\label{overlaptrace}
\ket{+}^{\otimes M} = \frac{1}{2^{M/2}} 
\sum_{\vec{x} \,\in\, \mathbb{F}_2^M} \ket{\vec{x}}
\end{equation}
and therefore
\begin{equation}
\label{traceform}
\bra{+}^{\otimes M}\ket{G}
= 2^{-M}\!\!
\sum_{\vec{x}\,\in\, \mathbb{F}_2^{M}}
\bra{\vec{x}} 
 \left(
 \prod_{i \sim j} CZ_{i,j}
 \right) 
 \ket{\vec{x}}
= 2^{-M}\,
\text{tr}\!
\left(\prod_{i \sim j} CZ_{i,j}\right).
\end{equation}
Since our calculation does not exploit any special properties of graph $G$, the calculation yields the trace of an \emph{arbitrary} product of $CZ$ gates---a quantity, which may be relevant independently of graph states.

\paragraph{Amplitude $\bra{+}^{\otimes M}\ket{G}$ as a sum over all induced subgraphs}
In Section~\ref{sec: Innerprod}, we motivated $\phi(\vec{x})$ by considering $\vec{x} \in \ker A$ and stabilizers $O_{\vec{x}}$. However, the characterization of $\phi(\vec{x})$ as counting edges in $H_{\vec{x}}$ applies to all $\vec{x} \in \mathbb{F}_2^M$. This has an interesting corollary. Now associating $\vec{x}$ with an eigenvector of $\otimes_i Z_i$, as we did above~(\ref{overlaptrace}), we observe that
\begin{equation}
    \label{czactiononx}
    \left(\prod_{i \sim j} CZ_{i,j}\right) \ket{\vec{x}} 
    = (-1)^{\phi(\vec{x})} \ket{\vec{x}}
\end{equation}
as a direct consequence of (\ref{defcz}). Here the quadratic form $\phi(\vec{x})$ is defined in equation~\eqref{phiexplicit1}:
\begin{equation}
    \phi(\vec{x})=\vec{x}^\intercal A^{\uppertr} \vec{x}. 
    \label{eq: def_phi}
\end{equation}
Then formula~(\ref{traceform}) gives:
\begin{equation}
    \bra{+}^{\otimes M}\ket{G} 
    = \frac{1}{2^M}\sum_{\vec{x}\,\in\, \mathbb{F}_2^M} (-1)^{\phi(\vec{x})}
    \label{evenminusodd}.
\end{equation}
In other words, we are counting induced subgraphs of $G$ with an even/odd number of edges and taking their difference. In the next subsection, we analyse the summation in (\ref{evenminusodd}) to determine the phase of $\bra{+}^{\otimes M}\ket{G}$.

\paragraph{Main result and organization} If $\bra{+}^{\otimes M}\ket{G}\neq 0$ is written in the form
  \begin{equation}
  \label{1stmainresult}
  \bra{+}^{\otimes M}\ket{G} = (-)^\Delta\, 2^{-(\text{\rm rank} A)/2}
  \end{equation}
then quantity~$\Delta \in \mathbb{F}_2$ is the Arf invariant of the quadratic refinement $\phi(\vec{x})=\vec{x}^\intercal A^{\uppertr} \vec{x}$ of the bilinear form $(\vec{x}, \vec{y}) \equiv \vec{x}^\intercal A \vec{y}$. 

In Subsection~\ref{sec:arf}, we briefly introduce the Arf invariant and explain that it agrees with the $\Delta$ in equation~\eqref{1stmainresult}. Subsection~\ref{sec: arf_interpret} interprets the calculation in terms of {\bf nonlocal effective qubits}. In Subsection~\ref{sec: arf_ex}, we calculate the Arf invariants (and symplectic bases) of prototypical graph states. 

The computations in this section can easily be extended to cover other partial amplitudes of $\ket{G}$ in the $X$-eigenbasis. This is done in Appendix~\ref{app:wfx}.

\subsection{Symplectic basis and the Arf invariant} 
\label{sec:arf}
The Arf invariant was first introduced by Arf in 1941 \cite{Arf1941}. A nice review of its mathematical properties is \cite{Dye1978}. Its most common application is in algebraic topology, where it partially characterizes spin structures of two-dimensional manifolds. In physics, it detects whether the Dirac operator admits an odd or even number of zero modes \cite{ASENS1971, Cimasoni:2006hmp,  Kapustin2015, Karch2019}. Here we explain that quantity $\Delta$ in (\ref{1stmainresult}) is the Arf invariant relevant to $A$ and $\phi$.

Motivated by equation~(\ref{evenminusodd}), study function $\phi(\vec{x})$. For arbitrary $\vec{x}, \vec{y} \in \mathbb{F}_2^M$, we have
\begin{align}
\phi(\vec{x} + \vec{y})
= (\vec{x}+\vec{y})^\intercal A^{\uppertr} (\vec{x} + \vec{y})
& = \phi(\vec{x}) + \phi(\vec{y})
  + \vec{x}^\intercal A^{\uppertr} \vec{y}
  + \vec{y}^\intercal A^{\uppertr} \vec{x} \\
& = \phi(\vec{x}) + \phi(\vec{y})
  + \vec{x}^\intercal A^{\uppertr} \vec{y}
  + \vec{x}^\intercal \big(A^{\uppertr}\big)^\intercal \vec{y}
\nonumber
    \label{appeqn:bilinear0}
\end{align}
and therefore:
\begin{equation}
    \phi(\vec{x} + \vec{y}) = \phi(\vec{x}) + \phi(\vec{y}) + \vec{x}^\intercal A \vec{y}
    \label{cohomology}
\end{equation}
This generalizes equation~(\ref{confirmhomom}), which was derived assuming $A\vec{y}=0$. In the study of bilinear forms over fields of characteristic 2, quantity $\phi(\vec{x})$ is referred to as a {\bf quadratic refinement} of the {\bf bilinear form}~$A\!: (\vec{x}, \vec{y}) \to \vec{x}^\intercal A \vec{y}$. To make progress, we must inspect the bilinear form more closely.

Let us decompose the vector space $\mathbb{F}_2^M$ into $\ker A \oplus U$.  
Assume $\ket{G} \neq \ket{+}^{\otimes M}$ so $U \neq \{0\}$. We wish to construct a \emph{symplectic} basis of $U$, in which the bilinear form $A$ takes a block diagonal form with blocks:
\begin{equation} \label{eq:sym=asym}
\begin{bmatrix} 0 & 1 \\ 1 & 0\end{bmatrix}
\equiv 
\begin{bmatrix} \phantom{-}0 & 1 \\ -1 & 0\end{bmatrix}.
\end{equation}
This is possible due to a standard result about antisymmetric matrices, which applies to $A$ because $A = A^\intercal \equiv -A^\intercal$ over $\mathbb{F}_2$. Note that this automatically implies that $\dim U = \text{\rm rank} A$ is even---a fact that can also be gleaned from equation~(\ref{1stmainresult}) when $\bra{+}^{\otimes M}\ket{G}\neq 0$. 

The symplectic basis can be found using an iterative procedure. To initiate it, we set $W_1 \equiv \ker A$ and $U_1 \equiv U$. Observe that all pairs $\vec{w} \in W_1$ and $\vec{u} \in U_1$ satisfy $\vec{w}^\intercal A \vec{u} = 0$ by the defining property of $W_1$. Then for every $1 \leq i \leq ({\rm rank}A)/2$ perform the following:
\begin{itemize}
    \item Given a vector $\vec{x}_i \in U_i$, find another vector $\vec{y}_i \in U_i$ such that $\vec{y}_i^\intercal A \vec{x}_i=1$. This is possible because $A\vec{x}_i \neq 0$ and at least one vector $\vec{z}_i \in \mathbb{F}_2^M$ is not orthogonal to $A\vec{x}_i$. (That vector cannot be $\vec{x}_i$ because $\vec{x}^\intercal A \vec{x}=0$ for all $\vec{x}$.) Such a $\vec{z}_i$ can be decomposed into $\vec{z}_i = \vec{w}_i + \vec{y}_i$ for some $\vec{w}_i \in W_i$ and $\vec{y}_i \in U_i$. Then $\vec{y}_i^\intercal A \vec{x}_i=1$ because $\vec{w}_i^\intercal A \vec{x}_i=0$.
    \item Set $W_{i+1} \equiv W_i \oplus {\rm span}\{\vec{x}_i, \vec{y}_i\}$ and $U_{i+1} = \{\vec{u}\in U_i\, |\,\vec{x}_i^\intercal A\vec{u}=\vec{y}_i^\intercal A\vec{u}=0\}$. Note that this definition maintains the condition $\vec{w}^\intercal A \vec{u} = 0$ for all pairs $\vec{w} \in W_{i+1}$ and $\vec{u} \in U_{i+1}$. Also note that $\dim U_{i+1} = \dim U_i - 2$ because $(\vec{x}_i - \vec{y}_i)^\intercal A \neq 0$.
    \item Iterate until $U_{i+1} = \{0\}$.
\end{itemize}
   
In this way, we have recast $\mathbb{F}_2^M$ as a direct sum
\begin{equation}
\label{directsumdecomposition}
    \{0,1\}^{\otimes M}=\ker A \oplus \left( \underset{i=1}{\overset{{\rm rank}A/2}{\bigoplus}} {\rm span}\left\{\vec{x}_i, \vec{y}_i\right\} \right),
\end{equation}
in which 
$\vec{y}_i^\intercal A \vec{x}_i=1$. The adjacency matrix $A$ in this basis (augmented by some basis $\vec{w}_k$ of $\ker A$) takes the canonical (anti-)symmetric form:
\begin{equation}
    V^\intercal A V
    =\left[ \begin{array}{ccc:cc:cc:cc}
0 & \cdots & 0 & & & & &  &  \\
\vdots & \ddots & \vdots &  & &  &  &  &  \\
0 & \cdots & 0 &  &  &  &  &  &  \\ \hdashline
 &  &  & 0 & 1 &  &  &  &  \\
 &  &  & 1 & 0 &  &  &  &  \\ \hdashline
 &  &  &  &  & 0 & 1 &  &  \\
 &  &  &  &  & 1 & 0 &  &  \\ \hdashline
 &  &  &  &  &  &  & ~\ddots \\
\end{array} \right] 
\label{appeqn: A_new}
\end{equation}
Here $V$ is the matrix whose columns are the basis vectors $\{\vec{w}_k\}_{k=1}^{\dim \ker A} \cup \{\vec{x}_i, \vec{y}_i\}_{i=1}^{({\rm rank} A)/2}$. The empty blocks in (\ref{appeqn: A_new}) are filled with zeros.

Let $\vec{p}, \vec{q}$ be any two vectors from this {\bf symplectic basis}.\footnote{Here we slightly abuse standard nomenclature by allowing $\ker A \neq \{0\}$ and including $\vec{w} \in \ker A$ in the `symplectic basis.' \label{symplecticbasisfootnote}} 
From equation~(\ref{cohomology}) and (\ref{appeqn: A_new}), we see that the behavior of $\phi$ is very rigid:
\begin{equation}
    \phi(\vec{p}+\vec{q})=
    \phi(\vec{p})+\phi(\vec{q})+
    \begin{cases}   
    1    &  {\rm if}~\vec{p}~{\rm and}~\vec{q}~{\rm are}~\vec{x}_i~{\rm and}~\vec{y}_i~\textrm{from the same block $i$} \\
    0    &  \text{otherwise}
    \end{cases}.
    \label{appeqn:phi_relate}
\end{equation}
Returning to equation~(\ref{evenminusodd}), we therefore obtain: 
\begin{align}
\label{allfactors}
    &
    \bra{+}^{\otimes M}\ket{G} 
    =
    \frac{1}{2^M}
    \sum_{\vec{x}\,\in\,\mathbb{F}_2^M}
    (-1)^{\phi(\vec{x})} \\
    =\, &
    \frac{1}{2^M} 
    \prod_{k=1}^{\dim \ker A} \Bigg(1 + (-1)^{\phi(\vec{w}_k)} \Bigg)
    \cdot\!\!
    \prod_{i=1}^{({\rm rank}A)/2}
    \Bigg(1 + (-1)^{\phi(\vec{x}_i)} + (-1)^{\phi(\vec{y}_i)} + (-1)^{1+\phi(\vec{x}_i)+\phi(\vec{y}_i)}\Bigg).
    \nonumber
\end{align}
The first product recovers Result~1 in Section~\ref{sec: Innerprod}---that the quantity vanishes if $\phi(\vec{w})=1$ for any $\vec{w} \in \ker A$. In what follows we assume this \emph{does not} happen.\footnote{At the end of this section, we explain that assuming $\phi(\vec{w})=0$ on $\ker A$ is necessary in order to define the Arf invariant---even if one studies the adjacency matrix $A$ for purposes other than computing $\bra{+}^{\otimes M}\ket{G}$. \label{deltaconsistent}}

We now inspect the second product in~(\ref{allfactors}). Up to a change of basis $\vec{x}_i\to \vec{x}_i+\vec{y}_i$ or $\vec{y}_i\to \vec{y}_i+\vec{x}_i$ in a block, the exponents in the parenthesis can only be: 
\begin{itemize}
    \item[($\flat$):] $\{\phi(\vec{x}_i), \phi(\vec{y}_i), \phi(\vec{x}_i) + \phi(\vec{y}_i) + 1\} =\{0,0,1\}$ and the factor in (\ref{allfactors}) evalues to $+2$. 
    \item[($\sharp$):] $\{\phi(\vec{x}_i), \phi(\vec{y}_i), \phi(\vec{x}_i) + \phi(\vec{y}_i) + 1\} =\{1,1,1\}$ and the factor in (\ref{allfactors}) evalues to $-2$.
\end{itemize}

All in all, equation~(\ref{allfactors}) has $\big(\!-M+(\dim \ker A) + ({\rm rank}A)/2\big) = -({\rm rank}A)/2$ factors of 2 and as many minus signs as there are blocks $i$ of type $(\sharp)$. Collecting everything altogether:
\begin{equation}
    \bra{+}^{\otimes{M}}\ket{G}=(-1)^\Delta \times 2^{-({\rm rank}A)/2}\times \prod_{i=1}^{\dim \ker A}\delta\big(\phi(\vec{w}_i)\big)
\end{equation}
Quantity~$\Delta$ counts (mod 2) blocks of type $(\sharp)$ in the symplectic basis of $A$. Because such blocks are distinguished by the condition $\phi(\vec{x}_i) = \phi(\vec{y}_i) = 1$, $\Delta$ can be written as:
\begin{equation}
    \Delta=\sum_{i=1}^{({\rm rank}A)/2} \phi(\vec{x}_i)\phi(\vec{y}_i) \pmod{2}.
\end{equation}
These facts \emph{define} the Arf invariant of the quadratic refinement~$\phi(\vec{x})$ of the bilinear form~$A\!: (\vec{x}, \vec{y}) \to \vec{x}^\intercal A \vec{y}$.

\subsection{Nonlocal effective qubits}  
\label{sec: arf_interpret}
In condensed matter theory, the Arf invariant detects gapped Majorana fermions in topological phases of matter \cite{Kapustin2015, Shiozaki_2017, Karch2019}. It is interesting to see what it detects in the context of stabilizer states.

\paragraph{Effective qubits}
The calculations in this section rely on expanding $\bra{+}^{\otimes M}$ and $\ket{G}$ in the $Z_1 \otimes Z_2 \otimes \ldots \otimes Z_M$ eigenbasis $\ket{\vec{x}}$, viz. equations~(\ref{overlaptrace}) and (\ref{czactiononx}). We now relabel this eigenbasis: 
\begin{equation}
\ket{\vec{x}} \equiv \underline{\ket{\vec{a}}}
\qquad {\rm where}~~
\vec{x} = V \vec{a}.
\label{atox}
\end{equation}
Here $V$ is the matrix with columns $\{\vec{w}_k\}_{k=1}^{\dim \ker A} \cup \{\vec{x}_i, \vec{y}_i\}_{i=1}^{({\rm rank} A)/2}$, which was used in equation~(\ref{appeqn: A_new}). Notice that relationship (\ref{atox}) is one-to-one because every $\vec{x}$ has a unique expansion
\begin{equation}
    \vec{x} = a_1\, \vec{w}_1 + a_2\, \vec{w}_2 + \cdots 
            + a_{M-1}\, \vec{x}_{({\rm rank}A)/2} + a_M\, \vec{y}_{({\rm rank}A)/2} \,,
            \label{xaexpansion}
\end{equation}
where the ellipses cover the sum over the full symplectic basis. We emphasize that in equation~(\ref{atox}) we are using a \emph{change of basis} in $\mathbb{F}_2^M$ (as a vector space under (mod 2) addition) to \emph{relabel a fixed basis} of the Hilbert space. Indeed, each $\underline{\ket{\vec{a}}}$ is still an eigenstate of $Z_1 \otimes Z_2\otimes \ldots \otimes Z_M$. 

We would like to think of the coefficients $a_i \in \mathbb{F}_2$ in expansion~(\ref{xaexpansion}) as defining $M$ {\bf effective qubits}, which are nonlocally realized by the $M$ {\bf physical qubits}. This means:
\begin{equation}
\vec{a} = \left(\,a_1\,a_2\, \ldots\, a_M\right)^\intercal
    \quad \Longleftrightarrow \quad
    \underline{\ket{\vec{a}}} = \ket{\underline{a_1}} \otimes 
    \underline{\ket{a_2}} \otimes
    \ldots \otimes
    \underline{\ket{a_M}}
    \label{whatisa}
\end{equation}
Each effective qubit is a representation of a single-qubit Pauli group, which is nonlocally embedded inside the $2^M$-dimensional Hilbert space of the $M$ physical qubits. Mimicking the usual representation of the Pauli group, we define \emph{effective $\underline{Z}_i$} and \emph{effective $\underline{X}_i$} operators:
\begin{align}
    \underline{Z}_i \underline{\ket{\vec{a}}} 
       & = (-)^{a_i} \underline{\ket{\vec{a}}} 
       \label{zona}\\
    \underline{X}_i \underline{\ket{\vec{a}}} 
       & = \underline{\ket{\vec{a} + \vec{e}_i}}
       \label{xona}
\end{align}
where $\vec{e}_i$ is the $i^{\rm th}$ unit vector in $a$-space. Because the basis $\underline{\ket{\vec{a}}}$ is the same as the $Z_1 \otimes Z_2\otimes \ldots \otimes Z_M$ eigenbasis $\ket{\vec{x}}$, only relabeled, it may be illuminating to exhibit action~(\ref{zona}-\ref{xona}) in the notation, which emphasizes the physical qubits:
\begin{align}
    \underline{Z}_i \ket{\vec{x}} 
       & = (-)^{(V^{-1}\vec{x})_i} \ket{\vec{x}} 
       \label{zonx}
       \\
    \underline{X}_i \ket{\vec{x}} 
       & = \begin{cases} 
       \ket{\vec{x} + \vec{w}_i}~~{\rm if}~1 \leq i \leq \dim \ker A\\
       \ket{\vec{x} + \vec{x}_j}~~{\rm if}~i - \dim\ker A = 2j -1\\
       \ket{\vec{x} + \vec{y}_j}~~{\rm if}~i - \dim\ker A = 2j
       \end{cases}
       \label{xonx}.
\end{align}
Equations~(\ref{zona}-\ref{xona}) and (\ref{zonx}-\ref{xonx}) are synonymous.

The Pauli algebra generated by $\underline{Z}_i$ and $\underline{X}_i$ implicitly defines the individual effective qubit states $\underline{\ket{0}}$ and $\underline{\ket{1}}$, which were invoked in equation~(\ref{whatisa}). 
By the same token, it also defines the $\underline{X}_i$ eigenstates
\begin{equation}
    \underline{\ket{+}} \equiv 
    \frac{1}{\sqrt{2}}
    \Big(\underline{\ket{0}} + \underline{\ket{1}}\Big)
    \qquad {\rm and} \qquad
    \underline{\ket{-}} \equiv 
    \frac{1}{\sqrt{2}}
    \Big(\underline{\ket{0}} - \underline{\ket{1}}\Big).
\end{equation}
in the Hilbert space of every effective qubit. For the purposes of preparing graph states, one attractive feature of $\underline{X}$-eigenstates is that:
\begin{equation}
    \ket{+}^{\otimes M} 
    = \frac{1}{2^{M/2}} \sum_{\vec{x}\,\in\, \mathbb{F}_2^M} \ket{\vec{x}} 
    = \frac{1}{2^{M/2}} \sum_{\vec{a}\,\in\, \mathbb{F}_2^M} \underline{\ket{\vec{a}}} 
    = \underline{\ket{+}}^{\otimes M}
    \label{equivalenceofpluses}.
\end{equation}
Notice that our calculation of $\bra{+}^{\otimes M}\ket{G}$ can be rewritten as:
\begin{equation}
\label{eq: re_trace}
    \bra{+}^{\otimes M}\ket{G}
= 2^{-M}\!\!
\sum_{\vec{x}\,\in\, \mathbb{F}_2^{M}}
\bra{\vec{x}} 
 \left(
 \prod_{i \sim j} CZ_{i,j}
 \right) 
 \ket{\vec{x}}
 = 2^{-M}\!\!
\sum_{\vec{a}\,\in\, \mathbb{F}_2^{M}}
\underline{\bra{\vec{a}}} 
 \left(
 \prod_{i \sim j} CZ_{i,j}
 \right) 
 \underline{\ket{\vec{a}}}.
\end{equation}
All our calculations can equally well be understood to have been carried out in the $a$-basis.

It is then interesting to understand how one would prepare the graph state $\ket{G}$ by directly manipulating the effective qubits, starting from the initial state $\underline{\ket{+}}^{\otimes M}$. In other words, what is the effective qubit analogue of the preparation of $\ket{G}$ described in equation~(\ref{eq:def_graph})?

\paragraph{Structure of graph states in terms of effective qubits} 
Combining equations~(\ref{czactiononx}) and (\ref{equivalenceofpluses}), we find:
\begin{equation}
\label{govera}
\ket{G} 
= 
\frac{1}{2^{M/2}} \sum_{\vec{x}\,\in\,\mathbb{F}_2^M} 
(-1)^{\phi(\vec{x})} \ket{\vec{x}}
=
\frac{1}{2^{M/2}} \sum_{\vec{a}\,\in\,\mathbb{F}_2^M} 
(-1)^{\underline{\phi}(\vec{a})} 
\underline{\ket{\vec{a}}},
\end{equation}
where $\underline{\phi}(\vec{a}) = \phi(V \vec{a})$.

Here we come to appreciate why it is useful to think in terms of the effective qubits. Equation~(\ref{appeqn:phi_relate}) says that the phase $\underline{\phi}(\vec{a})$ decomposes into a sum of $( \dim \ker A )$ single-qubit terms and $({\rm rank}A)/2$ two-qubit terms:
\begin{equation}
    \underline{\phi}(\vec{a}) 
  = \sum_{i = 1}^{\dim \ker A} \phi\big(a_i \cdot \vec{w}_i\big)
   + \sum_{j = 1}^{({\rm rank} A)/2} 
     \phi\Big(a_{(\dim \ker A)+2j-1} \cdot \vec{x}_j 
            + a_{(\dim \ker A)+2j} \cdot \vec{y}_j\Big).
\end{equation}
This means that $\ket{G}$ factorizes into a product of $( \dim \ker A )$ single effective qubit states and $({\rm rank}A)/2$ nonlocally realized `effective Bell pairs':
\begin{align}
    \ket{G} = 
    \bigotimes_{i=1}^{\dim \ker A} 
    \left( \frac{1}{\sqrt{2}}\,\underline{\ket{0}}
    +\frac{(-1)^{\phi(\vec{w}_i)}}{\sqrt{2}}\,\underline{\ket{1}}
    \right)
    \bigotimes_{j=1}^{({\rm rank}A)/2} 
    \Big( 
\underline{\ket{\,\flat\,}}~{\rm or}~\underline{\ket{\,\sharp\,}} 
    \Big)
    \label{gaexpansion}.
\end{align}

If any $\phi(\vec{w}_i) = 1$, the corresponding effective qubit is in the $\underline{\ket{-}}$ state and $\bra{+}^{\otimes M} \ket{G} = \underline{\bra{+}}^{\otimes M} \ket{G} = 0$. Let us recall how we characterized this circumstance in Section~\ref{sec: Innerprod}: $\bra{+}^{\otimes M} \ket{G}$ vanishes if and only if an induced subgraph $H$ exists such that every vertex in $G$ has an even number of neighbors in $H$ but $H$ has an odd number of edges. Evidently, this describes the preparation of effective $\underline{\ket{-}}$ states.

Assuming the product part of \eqref{gaexpansion} contains only $\underline{\ket{+}}$'s, the amplitude is determined by the effective Bell states. Their structure is fixed by the combination:
\begin{equation}
    \frac{1}{2} \left(
      \underline{\ket{00}} 
    +(-)^{\phi(\vec{x}_i)} \underline{\ket{10}} 
    +(-)^{\phi(\vec{y}_i)} \underline{\ket{01}} 
    +(-)^{\phi(\vec{y}_i)+\phi(\vec{x}_i)+1} \underline{\ket{11}}
    \right).
\end{equation}
Referring to the cases listed below equation \eqref{allfactors}, we find:
\begin{align}
    \underline{\ket{\,\flat\,}} 
    & = \tfrac{1}{2}\big( 
    \underline{\ket{00}} + \underline{\ket{10}} +
    \underline{\ket{01}} - \underline{\ket{11}}\big) 
    = \phantom{-}\underline{CZ} \,\underline{\ket{++}}, 
    \label{effbell1} \\
\underline{\ket{\,\sharp\,}} 
    & = \tfrac{1}{2}\big( 
    \underline{\ket{00}} - \underline{\ket{10}} -
    \underline{\ket{01}} - \underline{\ket{11}} \big)
    = - (\underline{X} \otimes \underline{X}) \,\underline{CZ}\,
    (\underline{X} \otimes \underline{X}) \,\underline{\ket{++}}
    \label{effbell2}.
\end{align}

These Bell pairs call for a few explanatory comments:
\begin{itemize}
\item In operational terms, the effective Bell pair indicated in (\ref{effbell1}) is unambiguously defined. Yet the notation with which we express it depends on the choice of the symplectic block basis $\{\vec{x}_i, \vec{y}_i\}$ assumed in equations~(\ref{zonx}-\ref{xonx}). For example, had we worked with the tilded basis $\vec{\tilde{x}}_i = \vec{x}_i$ and $\vec{\tilde{y}}_i = \vec{y}_i + \vec{x}_i$, the \emph{same} effective Bell pair would have been denoted:
\begin{equation}
\underline{\ket{\,\tilde{\flat}\,}} 
    = \tfrac{1}{2}\big( 
    \underline{\ket{\tilde{0}\tilde{0}}} + \underline{\ket{\tilde{1}\tilde{0}}} +
    \underline{\ket{\tilde{1}\tilde{1}}} - \underline{\ket{\tilde{0}\tilde{1}}} \big)
\label{effbell1b}
\end{equation}
\item Conjugation by $\underline{X}\otimes \underline{X}$ in the definition of $\underline{\ket{\,\sharp\,}}$ is immaterial for computing $\bra{+}^{\otimes M} \ket{G}$ because the latter quantity is a trace, viz. equation~\eqref{traceform}. The $\underline{X}\otimes \underline{X}$ cancel out by the cyclicity of the trace.
\item On the other hand, the minus sign in \eqref{effbell2} does affect $\bra{+}^{\otimes M} \ket{G}$. Referring back to \eqref{gaexpansion}, we see that the sign of $\bra{+}^{\otimes M} \ket{G}$ arises from counting (mod 2) the effective Bell pairs of the $(\sharp)$ type. Evidently, this is what the Arf invariant $\Delta$ does.
\end{itemize}

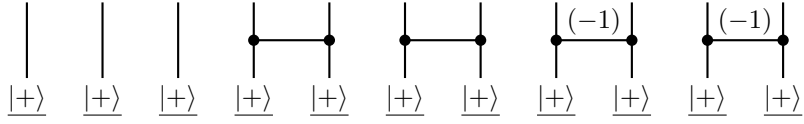
\begin{figure}[tbp]
    \centering
    \begin{tikzpicture}[
    vertical_line/.style = {thick},
    solid_connector/.style = {thick},
    dashed_connector/.style = {dashed, thick},
    intersection_point/.style = {fill, circle, inner sep=1.5pt}
]

\foreach \i in {1, 2, 3} {
    \node[draw=none] (l\i) at (\i, 0) {};
    \draw[vertical_line] (l\i.center) -- ++(0, 1);
}

\foreach \i in {1, 2, ..., 4} {
    \node[draw=none] (m\i) at (3 + \i, 0) {};
    \draw[vertical_line] (m\i.center) -- ++(0, 1);
}

\foreach \i in {1, 2, ..., 4} {
    \node[draw=none] (r\i) at (7 + \i, 0) {};
    \draw[vertical_line] (r\i.center) -- ++(0, 1);
}

\foreach \i/\j in {1/2, 3/4} {
    \coordinate (start) at ($(m\i.center) + (0, 0.5)$);
    \coordinate (end) at ($(m\j.center) + (0, 0.5)$);
    \draw[solid_connector] (start) -- (end);
    \node[intersection_point] at (start) {};
    \node[intersection_point] at (end) {};
}

\foreach \i/\j in {1/2, 3/4} {
    \coordinate (start) at ($(r\i.center) + (0, 0.5)$);
    \coordinate (end) at ($(r\j.center) + (0, 0.5)$);
    \draw[solid_connector] (start) -- (end) node[midway, above, yshift=-2pt] {\small$(-1)$};
    \node[intersection_point] at (start) {};
    \node[intersection_point] at (end) {};
}

\foreach \i in {1, 2, ..., 11} {
    \node at (\i, -0.3) {\small$\underline{\ket{+}}$};
}

\end{tikzpicture}
\caption{In the basis of effective qubits, every graph state with non-vanishing $\braket{+|^{\otimes M} | G}$ amplitude takes this form. When $\braket{+|^{\otimes M} | G}=0$, the analogous decomposition includes both $\underline{\ket{+}}$ and $\underline{\ket{-}}$ states.}
    \label{appfig:new_graph}
\end{figure}

In summary, we may view the preparation of a graph state $\ket{G}$ with $\bra{+}^{\otimes M} \ket{G} \neq 0$ as being implemented by the circuit shown in Figure~\ref{appfig:new_graph}.  The Arf invariant $\Delta$ counts the phase factors in this circuit.

\paragraph{Arf invariant assumes no $\underline{\ket{-}}$ states}
In footnote~\ref{deltaconsistent} we remarked that defining the Arf invariant relies on $\phi(\vec{w})=0$ for every $\vec{w} \in \ker A$. In other words, quantity $\Delta$ is well defined only if state~(\ref{gaexpansion}) contains no $\underline{\ket{-}}$. Let us see why.

Suppose that the symplectic basis of $A$ contains three vectors $\vec{w}$, $\vec{x}$ and $\vec{y}$ such that
\begin{align}
    A\vec{w} = 0 \qquad &{\rm and} \qquad \phi(\vec{w}) = 1 \nonumber \\
    A\vec{x} \neq 0 \qquad &{\rm and} \qquad \phi(\vec{x}) = 0 \\
    A\vec{y} \neq 0 \qquad &{\rm and} \qquad \phi(\vec{y}) = 0 
    \qquad {\rm and} \qquad \phi(\vec{x} + \vec{y}) = 1 \nonumber
\end{align}
Their being part of the symplectic basis also means $\phi(\vec{w}+\vec{x}) = \phi(\vec{w}) + \phi(\vec{x}) = 0$ and likewise for $\vec{y}$. This is a setup, in which these three vectors contribute $\underline{\ket{-}} \otimes \underline{\ket{\,\flat\,}}$ to decomposition (\ref{gaexpansion}). However, we may also choose another symplectic basis
\begin{equation}
    \vec{x}\,' = \vec{x} + \vec{w} \qquad {\rm and} \qquad
    \vec{y}\,' = \vec{y} + \vec{w}\,,
\end{equation}
which satisfies:
\begin{align}
    \phi(\vec{x}\,') & = \phi(\vec{x}) + \phi(\vec{w}) = 1 \nonumber ,\\
    \phi(\vec{y}\,') & = \phi(\vec{y}) + \phi(\vec{w}) = 1 ,\\
    \phi(\vec{x}\,' + \vec{y}\,') & = \phi(\vec{x} + \vec{y}) = 1.
\end{align}
In the new symplectic basis, $\{\vec{w}, \vec{x}\,', \vec{y}\,'\}$ contribute $\underline{\ket{-}} \otimes \underline{\ket{\,\sharp\,}}$ to $\ket{G}$. Thus, the presence of a $\underline{\ket{-}}$ allows us to freely convert $\underline{\ket{\,\flat\,}} \leftrightarrow \underline{\ket{\,\sharp\,}}$.

\subsection{Examples} 
\label{sec: arf_ex}
As illustration, we find symplectic bases and compute the Arf invariants of one-dimensional cluster states (chains and rings) and graph states based on star graphs and complete graphs. As elsewhere, $M$ denotes the number of qubits in the graph state.

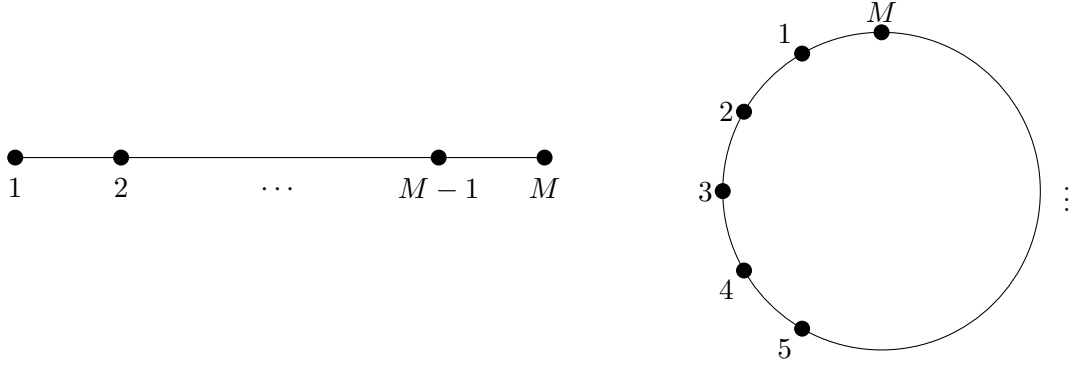
\begin{figure}[tbp]
\centering

\begin{minipage}{0.45\textwidth}
\centering
\begin{tikzpicture}[scale=1.4]
    \draw (-2.5,0) -- (2.5,0);
    \foreach \x in {-2.5,-1.5,1.5,2.5}
        \filldraw (\x,0) circle (2pt);
    \node at (0,-0.3) {$\cdots$};
    \node[below] at (-2.5,-0.1) {$1$};
    \node[below] at (-1.5,-0.1) {$2$};
    \node[below] at (1.5,-0.1) {$M-1$};
    \node[below] at (2.5,-0.1) {$M$};
\end{tikzpicture}
\end{minipage}
\hfill
\begin{minipage}{0.45\textwidth}
\centering
\begin{tikzpicture}[scale=1.4]
    \draw (0,0) circle (1.5cm);
    
    \foreach \angle in {90,120,150,180,210,240}
        \filldraw (\angle:1.5) circle (2pt);
    
    \node at (0:1.75) {$\vdots$};
    
     \node[above] at (90:1.5) {$M$};
    \node[above left] at (120:1.5) {$1$};
    \node[left] at (150:1.5) {$2$};
    \node[left] at (180:1.5) {$3$};
    \node[below left] at (210:1.5) {$4$};
    \node[below left] at (240:1.5) {$5$};
\end{tikzpicture}
\end{minipage}
\caption{One-dimensional cluster states: chains (left) and rings (right).}
\label{fig:cluster_graph}
\end{figure}

\paragraph{Chain cluster states} 
The underlying graph is shown in the left panel of Figure~\ref{fig:cluster_graph}. These states are often discussed in the context of Measurement-Based Quantum Computation (MBQC) because they realize the so-called quantum wire \cite{PhysRevLett.108.240505}. This means that by measuring the chain cluster state, one can emulate a general single-qubit unitary transformation. A single run of such a measurement-based computation returns a single bit of output. 

The dimension of $\ker A$ depends on the parity of $M$. For odd $M$, $\ker A$ is spanned by $\{\vec{w}_1=(1,0,1,0,\dots ,0,1)^\intercal \}$. For even $M$, $\dim \ker A = 0$. One valid set of symplectic blocks (often called \emph{hyperbolic blocks}) $\{\vec{x}_i, \vec{y}_i\}$
is:
\begin{equation}
    (\vec{x}_i)_j=\begin{cases}
    1 & j \text{ odd},~j<2i \\
    0 & \text{others} \\
    \end{cases} \qquad
    (\vec{y}_i)_j=\delta_{j,2i}
    \qquad 
    \Big(1\leq i\leq \left\lfloor M/2 \right \rfloor;~~1\leq j\leq M \Big)
    \label{eqn:sym_base_cluster}
\end{equation}
Thus, matrices~$V$ evoked in equation~(\ref{appeqn: A_new}) are:
\begin{equation}
\begin{array}{cp{1cm}c}
        V_{\textrm{($M$~odd)}} =
        \left[ \begin{array}{c:cc:cc:cc:c}
     1~ & ~1 & 0 & 1 & 0 & 1 & 0 &  \\
     0~ & ~0 & 1 & 0 & 0 & 0 & 0 &  \\
     \hdashline
     1~ & ~0 & 0 & 1 & 0 & 1 & 0 &  \\
     0~ & ~0 & 0 & 0 & 1 & 0 & 0 &  \\
     \hdashline
     1~ & ~0 & 0 & 0 & 0 & 1 & 0 &  \\
     0~ & ~0 & 0 & 0 & 0 & 0 & 1 &  \\
     \hdashline
     \vdots~ & & & & & & & ~\ddots
    \end{array} \right]
& &
        V_{\textrm{($M$~even)}} =
\left[ \begin{array}{cc:cc:cc:c}
    1 & 0 & 1 & 0 & 1 & 0 &  \\
    0 & 1 & 0 & 0 & 0 & 0 &  \\
    \hdashline
    0 & 0 & 1 & 0 & 1 & 0 & \\
    0 & 0 & 0 & 1 & 0 & 0 &  \\
    \hdashline
    0 & 0 & 0 & 0 & 1 & 0 &  \\
    0 & 0 & 0 & 0 & 0 & 1 &  \\
    \hdashline
    & & & & & & ~\ddots
    \end{array} \right]
    \end{array}
\end{equation}
Since subgraphs $H_{\vec{x}_i}$ and $H_{\vec{y}_i}$ contain no edges, $\phi(\vec{x}_i)=\phi(\vec{y}_i)=0$. Therefore all the blocks are of type ($\flat$) and the Arf invariants of all chain cluster states are $\Delta = 0$.

\paragraph{Ring cluster states}
The underlying graph is shown in the right panel of Figure~\ref{fig:cluster_graph}. Ring cluster states have been used to show that computational outcomes in Measurement-Based Quantum Computation are fluxes of a $\mathbb{Z}_2 \times \mathbb{Z}_2$ gauge theory \cite{Wong_2024}. Each MBQC run on an even-membered ring cluster state yields two classical bits as computational outcomes. 

For odd $M$, $\ker A$ is spanned by $\vec{w}_1=(1,1,\dots,1,1)^\intercal$. In this case, $\phi(\vec{w}_1)=1$ because the graph contains an odd number of edges. Consequently, the Arf invariant is not defined. 

For even $M$, $\ker A$ is spanned by $\vec{w}_1=(1,0,1,0,\dots,1,0)^\intercal$ and $\vec{w}_2=(0,1,0,1,\dots,0,1)^\intercal$. Now $\phi(\vec{w}_1)=\phi(\vec{w}_2)=0$ so the Arf invariant is well defined. 

Equation \eqref{eqn:sym_base_cluster} still gives a valid set of symplectic blocks, except that there is one fewer of them for even $M$: 
\begin{equation}
    (\vec{x}_i)_j=\begin{cases}
    1 & j \text{ odd},~j<2i \\
    0 & \text{others} \\
    \end{cases} \qquad
    (\vec{y}_i)_j=\delta_{j,2i}
    \qquad 
    \Big(1\leq i\leq \left\lfloor (M-1)/2 \right \rfloor;~~1\leq j\leq M \Big)
    \label{eqn:sym_base_cluster2}
\end{equation}
For even $M$, we must inspect the symplectic blocks to determine the Arf invariant. Once again, all subgraphs $H_{\vec{x}_i}$ and $H_{\vec{y}_i}$ contain no edges so the blocks are all of type ($\flat$) and $\Delta = 0$.

\begin{figure}[tbp]
\centering

\begin{minipage}{0.45\textwidth}
\centering
\begin{tikzpicture}[scale=1.1]
    \filldraw (0,0) circle (2pt) node[above] {$1$};
    \foreach \angle in {0,45,135,180,225,270,315}
        \filldraw (\angle:2) circle (2pt);
    \node at (90:2) {$\cdots$};
    
    \foreach \angle in {0,45,135,180,225,270,315}
        \draw (0,0) -- (\angle:2);
    
    \node[above right] at (0:2) {$2$};
    \node[below right] at (315:2) {$3$};
    \node[below left] at (225:2) {$5$};
    \node[below] at (270:2) {$4$};
    \node[above left] at (180:2) {$6$};
    \node[above left] at (135:2) {$7$};
    \node[above right] at (45:2) {$M$};

\end{tikzpicture}
\end{minipage}
\hfill
\begin{minipage}{0.45\textwidth}
\centering
\begin{tikzpicture}[scale=1.65]
    \foreach \i in {1,...,5} {
        \node[circle, fill, inner sep=2pt] (v\i) at ({\i*72+18}:1.5) {};
        \node at ({\i*72+18}:1.8) {$\i $};
    }
    
    \foreach \i in {1,...,5} {
        \foreach \j in {\i,...,5} {
            \ifnum\i=\j
            \else
                \draw (v\i) -- (v\j);
            \fi
        }
    }
\end{tikzpicture}
\end{minipage}
\caption{A star graph (left) and a complete graph (right; here shown for $M=5$). Both types of graph states are locally Clifford-equivalent to GHZ states.}
\label{fig:star_graph}
\end{figure}
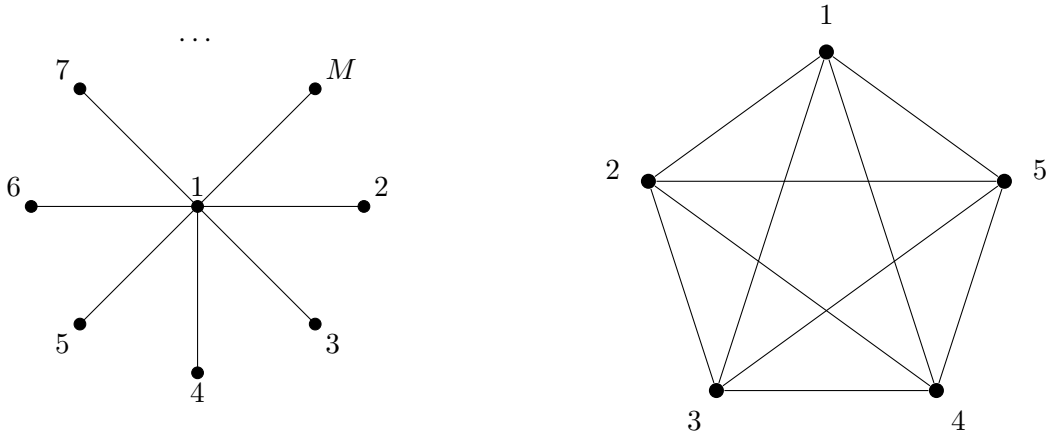

\paragraph{Star graph states} 
The underlying graph is displayed on the left of Figure~\ref{fig:star_graph}. We have $\dim \ker A = M-2$, spanned by $\{\vec{w}_1=(0,1,1,0,0,\dots)^\intercal$, $\vec{w}_2=(0,1,0,1,0,\dots)^\intercal$, $\vec{w}_3=(0,1,0,0,1,\dots)^\intercal,\dots \}$. With ${\rm rank}A=2$, we have a single symplectic block, which can be chosen as $\{\vec{x}_1=(1,0,0,\dots)^\intercal,\vec{y}_1=(0,1,0,\dots )^\intercal\}$. We see that
\begin{equation}
\phi(\vec{w}_1)=\phi(\vec{w}_2)=\dots=\phi(\vec{x}_1)=\phi(\vec{y}_1)=0
\end{equation}
and therefore $\Delta=0$.

An important note that is the star graph state is locally Clifford (LC)-equivalent to the GHZ state (on $M$ qubits) \cite{PhysRevA.69.022316}. Notice, however, that while
\begin{equation}
\braket{+|^{\otimes M} | G} = 2^{-({\rm rank}A)/2} = 1/2
\end{equation}
we have
\begin{equation}
\braket{+|^{\otimes M} | {\rm GHZ}} = 
2^{-M/2}\!\!
\sum_{\vec{x}\,\in\, \mathbb{F}_2^{M}}
\bra{\vec{x}} 
\left(\frac{\ket{000\ldots}}{\sqrt{2}} + \frac{\ket{111\ldots}}{\sqrt{2}}\right)
=
2^{-(M-1)/2}.
\end{equation}
This is of course not surprising because state $\bra{+}^{\otimes M}$ is not LC-invariant. 

For a general stabilizer state $\ket{\Psi}$, the inner product $\braket{+|^{\otimes M} | \Psi}$ can be computed in our formalism by bringing $\ket{\Psi}$ to a graph state form using a local Clifford transformation \emph{and}, at the same time, transforming $\bra{+}^{\otimes M}$ by the inverse Clifford transformation. The resulting object is generally a known linear combination of the $2^M$ amplitudes $\braket{\pm \pm \ldots | G}$, which are computed in Appendix~\ref{app:wfx}.

\paragraph{Complete graph states} 
A complete graph is shown on the right of Figure~\ref{fig:star_graph}. 

For $M$ odd $\ker A$ is spanned by $\vec{w}_1=(1,1,1,\dots,1)^\intercal$ while for $M$ even $\dim \ker A = 0$. The following choice of symplectic blocks works for all $M$:
\begin{equation}
    (\vec{x}_i)_j=\begin{cases}
    1 & j\leq 2i-1 \\
    0 & \text{others} \\
    \end{cases} \quad
    (\vec{y}_i)_j=\begin{cases}
    1 & j\leq 2i-2 \text{ or } j=2i \\
    0 & \text{others} \\
    \end{cases}
        \qquad 
    \Big(1\leq i\leq \left\lfloor M/2 \right \rfloor \Big)
    \label{eqn:sym_base_compl}
\end{equation}
More explicitly, the matrices $V$ referenced in equation~(\ref{appeqn: A_new}), with columns written in the order $\vec{w}_1$ (for $M$ odd) then $\vec{x}_1, \vec{y}_1,\vec{x}_2,\vec{y}_2,\dots$ are:
\begin{equation}
\begin{array}{cp{1cm}c}
        V_{\textrm{($M$~odd)}} =
        \left[ \begin{array}{c:cc:cc:cc:c}
     1~ & ~1 & 0 & 1 & 1 & 1 & 1 &  \\
     1~ & ~0 & 1 & 1 & 1 & 1 & 1 &  \\
     \hdashline
     1~ & ~0 & 0 & 1 & 0 & 1 & 1 &  \\
     1~ & ~0 & 0 & 0 & 1 & 1 & 1 &  \\
     \hdashline
     1~ & ~0 & 0 & 0 & 0 & 1 & 0 &  \\
     1~ & ~0 & 0 & 0 & 0 & 0 & 1 &  \\
     \hdashline
     \vdots~ & & & & & & & ~\ddots
    \end{array} \right]
& &
        V_{\textrm{($M$~even)}} =
\left[ \begin{array}{cc:cc:cc:c}
    1 & 0 & 1 & 1 & 1 & 1 &  \\
    0 & 1 & 1 & 1 & 1 & 1 &  \\
    \hdashline
    0 & 0 & 1 & 0 & 1 & 1 & \\
    0 & 0 & 0 & 1 & 1 & 1 &  \\
    \hdashline
    0 & 0 & 0 & 0 & 1 & 0 &  \\
    0 & 0 & 0 & 0 & 0 & 1 &  \\
    \hdashline
    & & & & & & ~\ddots
    \end{array} \right]
    \end{array}
\end{equation}
Counting edges, we find:
\begin{itemize}
    \item $\phi(\vec{x}_i)=\phi(\vec{y}_i)=0$ when $i$ is odd. This block is of type ($\flat$).
    \item $\phi(\vec{x}_i)=\phi(\vec{y}_i)=1$ when $i$ is even. This block is of type ($\sharp$).
\end{itemize}
Thus, \emph{in this basis} we have an even number of $(\sharp)$-type blocks when $M \equiv 0, 1, 2, 3~({\rm mod}~8)$ and an odd number when $M \equiv 4, 5, 6, 7~({\rm mod}~8)$. For even $M$, this fully determines the Arf invariant because no vectors from $\ker A$ can render $\Delta$ ill-defined. In contrast, for odd $M$ we must inspect:
\begin{equation}
\phi(\vec{w}_1) \equiv \frac{M(M-1)}{2}~({\rm mod}~2) =
\begin{cases} 0 & ~~M \equiv 0,1~~({\rm mod}~4) \\ 1 & ~~M \equiv 2,3~~({\rm mod}~4) \end{cases}.
\end{equation}
Consequently, the Arf invariant is not defined for $M \equiv 3, 7~({\rm mod}~8)$. In summary:
\begin{equation}
    \Delta = 
    \begin{cases}
    0 & M\equiv 0,1,2~({\rm mod}~8) \\
    1 & M\equiv 4,5,6~({\rm mod}~8) \\
    \text{not defined} & M \equiv 3,7~({\rm mod}~8)
    \end{cases}
\end{equation}

Complete graph states are \emph{also} LC-equivalent to GHZ states \cite{PhysRevA.69.022316}. We reiterate the point that the inner product $\braket{+|^{\otimes M} | \Psi}$ is not LC-invariant \emph{and} add a new observation: Since star graph states and complete graph states are LC-equivalent to one another, we see that local Clifford transformations also affect the Arf invariant.

\section{Expectation value of a permutation} 
\label{sec:compute_permut}
In the preceding sections we computed the inner product $\braket{G' | G''}$. In this section, we compute the same quantity under an additional assumption: that $\ket{G''} = U(\pi) \ket{G'}$, where $\pi \in S_M$ permutes the constituent qubits. This circumstance is particularly relevant to computing multi-invariants \cite{Gadde2025}. For computations of multi-invariants of graph states, see the recent paper \cite{akella2026multiinvariantsstabilizerstates} and Appendix~\ref{appendix:MI}. 

In what follows we drop the primes and compute:
\begin{equation}
\label{eq:trpiG}
    \bra{G}U(\pi)\ket{G}={\rm tr}\, U(\pi)\ket{G}\bra{G}
\end{equation}
The action of the permutation $\pi$ on the Hilbert space
\begin{equation}\label{eq:permconvention}
    U(\pi)\ket{x_1 x_2 \dots x_M}=\ket{x_{\pi(1)} x_{\pi(2)}\dots x_{\pi(M)}}
\end{equation}
must be distinguished from its action on the vector space $\mathbb{F}_2^M$:
\begin{equation}
    (P\vec{x})_i=(\vec{x})_{\pi^{-1}(i)}\,,
    \qquad {\rm where}~~\vec{x}=(x_1\,\ldots\,x_M)^\intercal \in \mathbb{F}_2^M
    \label{defP}
\end{equation}
With these conventions, we have $U(\pi)\ket{\vec{x}}=\ket{P \vec{x}}$. 

Naturally, one way to compute $\bra{G}U(\pi)\ket{G}$ is to follow the route of Sections~\ref{sec: Innerprod} and \ref{sec: Arf_inv}. If the adjacency matrix of $G$ is $A$ then the permuted state $U(\pi)\ket{G}$ is a graph state with adjacency matrix $P^\intercal \! A P$ and therefore: 
\begin{equation}
    \bra{G}U(\pi)\ket{G}
    =\frac{1}{2^M}\sum_{\vec{x}\,\in\,\mathbb{F}_2^M}(-1)^{\vec{x}^\intercal (A+P^\intercal\! A P)^{\suppertr} \vec{x}} 
    = \pm\, 2^{-{\rm rank}(A+P^\intercal\! A P)/2}
    \label{oldapproach}
\end{equation}
In effect, we have simply replaced $A \to A+P^\intercal\! A P$. However, it turns out that when the adjacency matrix has the special structure $A+P^\intercal \! A P$, a more customized approach becomes viable.  

We present the alternative approach in Section~\ref{subsec:tracecycles} and compare it with equation~(\ref{oldapproach}) in Section~\ref{subsec:comparetwoapproaches}. The approach in this section is often more suitable when the permutation is determined by a few tunable parameters, as is the case in multi-invariant calculations \cite{Gadde_2022,Penington_2023,Harper_2024,Gadde2025,Harper:2025uui}. We illustrate this in Appendix~\ref{appendix:MI}, which computes multi-invariants \cite{Gadde2025} of several commonly encountered classes of stabilizer states.

\subsection{Customized approach to expectation values of permutations}
\label{subsec:tracecycles}
From equation~(\ref{eq:rho_stab}), the density matrix (outer product) of a graph state is expressed in terms of the stabilizer group generators $K_i$ (equation~\ref{KintermsofA}) as:
\begin{equation}
\label{allstabs}
    \ket{G}\bra{G}
    =\frac{1}{2^M}\sum_{\vec{x}\,\in\,\mathbb{F}_2^M}O_{\vec{x}}
    =\frac{1}{2^M}\sum_{\vec{x}\,\in\,\mathbb{F}_2^M} K_1^{x_1} K_2^{x_2} \dots K_M^{x_M}
\end{equation}
The stabilizers $O_{\vec{x}}$ are explicitly given by formula~(\ref{arbitrarystab}) 
\begin{equation}
    O_{\vec{x}} =
    (-)^{\phi(\vec{x})} \cdot X_1^{x_1} Z_1^{(A\vec{x})_1} \otimes X_2^{x_2} Z_2^{(A\vec{x})_2} \otimes \ldots \otimes X_M^{x_M} Z_M^{(A\vec{x})_M}
    \label{arbstabexact}
\end{equation}
with the sign computed by equation~(\ref{phiexplicit1}). We observed this before in equation~(\ref{kernelstab}) in the special case $A\vec{x} = 0$, but in fact (\ref{arbstabexact}) works for all $\vec{x}\in\mathbb{F}_2^M$. The proof is left as an exercise for the reader.

Suppose permutation $\pi$ decomposes into $r$ cycles $C_s$ of length $l_s$ ($1\leq s\leq r$):
\begin{equation}
\label{eq: perm_conv}
    \pi=C_1\, C_2 \cdots C_r
    =(1\, 2\,\dots\, l_1)(l_1+1\,\,\, l_1+2\,\dots\, l_1+l_2)\cdots(M-l_r+1\,\dots\, M).
\end{equation}
The second equality assumes that the vertices of $G$ are appropriately indexed. In what follows we use the shorthand $i \in C_s$ to express that the $s^{\rm th}$ cycle of $\pi$ acts on the $i^{\rm th}$ vertex of $G$: 
\begin{equation}
 i \in C_s: \qquad   l_1+\dots+l_{s-1}+1\leq i \leq l_1+\dots+l_{s}.
\end{equation}

From equation~(\ref{allstabs}), expression~(\ref{eq:trpiG}) decomposes into a sum over stabilizers $O_{\vec{x}}$. From the cycle decomposition, each summand is a product over cycles. All in all, we find:
\begin{align}
    {\rm tr}\big(U(\pi)\ket{G}\bra{G}\big)
    &=\frac{1}{2^M} \sum_{\vec{x}\,\in\,\mathbb{F}_2^M} {\rm tr}\,U(\pi)\,O_{\vec{x}} \notag \\
    &=\frac{1}{2^M}\sum_{\vec{x}\,\in\,\mathbb{F}_2^M}
    (-)^{\phi(\vec{x})}\prod_{s=1}^r 
    {\rm tr}\left[
        U(C_s)\,\, \underset{i\in C_s}{\bigotimes}\!\left(X_i^{x_i}Z_i^{(A\vec{x})_i}\right)\right]
    \label{trpiG_expanded}
\end{align}
The remaining work is to evaluate 
\begin{align}
    &
    {\rm tr}\,
        U(C_s)\, \underset{i\in C_s}{\bigotimes}\!\left(X_i^{x_i}Z_i^{(A\vec{x})_i}\right)
    \notag \\
    =\,&{\rm tr}\,
        U\Big((12\dots l_s)\Big)
            \left(X_1^{a_1} Z_1^{b_1}\right)\otimes
            \left(X_2^{a_2} Z_2^{b_2}\right)\otimes
            \cdots \otimes 
            \left(X_{l_s}^{a_{l_s}} Z_{l_s}^{b_{l_s}}\right)
\end{align}
for each cycle $C_s$, with $\vec{a}, \vec{b}\in\{0,1\}^{\otimes\, l_s}$ playing the role of $\vec{x}$ and $A\vec{x}$ restricted to $i \in C_s$. With convention~\eqref{eq:permconvention}, this expression evaluates to: 
\begin{align}
    &\sum_{x_1\ldots\,x_{l_s}=0,1} 
    \bra{x_1 x_2 \dots x_{l_s}}
    \overset{l_s}{\underset{i=1}{\bigotimes}}\left(X_i^{a_i}Z_i^{b_i}\right)
    \ket{x_2\dots x_{l_s} x_1} \notag\\
    =\,&\sum_{x_1\ldots\, x_{l_s}=0,1} 
    \bra{x_1}X^{a_1}Z^{b_1}\ket{x_2}\bra{x_2}X^{a_2}Z^{b_2}\ket{x_3}\bra{x_3}\cdots \ket{x_{l_s}}\bra{x_{l_s}}X^{a_{l_s}}Z^{b_{l_s}}\ket{x_1} \notag\\
    =\,&\text{tr}\,X^{a_1}Z^{b_1}X^{a_2}Z^{b_2}\cdots X^{a_{l_s}}Z^{b_{l_s}}.
\end{align}
In the second and third line the Pauli matrices carry no subscripts because the expectation values / trace are taken in a single-qubit Hilbert space.

Because $X$, $Z$ and $XZ$ are traceless, this expression is only non-zero when both $\sum_{i=1}^{l_s} a_i$ and $\sum_{i=1}^{l_s} b_i$ are even, in which case it equals $\text{tr}(\pm I) = \pm 2$. The sign comes from $XZ=-ZX$. By commuting all the $X$s past the $Z$s to the left, we recognize that the number of $(-)$s is $\sum_{i>j}a_i b_j$.\footnote{Another way to determine the sign is to commute all the $X$'s to the right past the $Z$s. This gives $\sum_{i\leq j}a_i b_j$, whose parity is the same as the parity of $\sum_{i> j}a_i b_j$ when the trace is nonvanishing because $\sum_{i\leq j}a_ib_j+\sum_{i>j}a_ib_j=(\sum_{i}a_i)(\sum_{j}b_j)$ is even.} In conclusion, we find:
\begin{align}
&{\rm tr}\,
        U\Big((12\dots l_s)\Big)
            \left(X_1^{a_1} Z_1^{b_1}\right)\otimes
            \left(X_2^{a_2} Z_2^{b_2}\right)\otimes
            \cdots \otimes 
            \left(X_{l_s}^{a_{l_s}} Z_{l_s}^{b_{l_s}}\right) 
            \notag \\
    =\,&2\times (-)^{\sum_{i>j}a_i b_j} 
    \times \delta_{\sum_{i=1}^{l_s} a_i,\,{\rm even}}\, \delta_{\sum_{i=1}^{l_s} b_i,\,{\rm even}}
    \label{tracesinglecycle}
\end{align}

Substituting this back in~\eqref{trpiG_expanded}, we finally obtain a customized formula for the expectation value of a permutation in a graph state:
\begin{align}
    &\bra{G}U(\pi)\ket{G}
    \notag \\
    &=\frac{2^r}{2^M}\sum_{\vec{x}\,\in\, \mathbb{F}_2^M}
        (-)^{\phi(\vec{x})}\, \prod_{s=1}^r \,
        \left[ (-)^{\sum_{(i>j)\in C_s}(\vec{x})_i (A\vec{x})_j}
        \times \delta_{\,\sum_{i \in C_s} (\vec{x})_i,\,{\rm even}}\, \delta_{\,\sum_{j \in C_s} (A \vec{x})_j,\,{\rm even}}
        \right]
    \label{eq:trpiG_result}
\end{align}
We remind the reader that the exponent assumes that the indexing of the qubits respects the ordering of the qubits within the cycles $C_s$.

For later convenience, we subject the Kronecker deltas to a further rewriting. For a fixed $\pi$ given by equation~(\ref{eq: perm_conv}), we introduce a family of $r$ vectors $\vec{p}_s$
\begin{equation}
\label{defp}
    (\vec{p}_s)_i=\begin{cases}
    1 & {\rm if}~i\in C_s \\
    0 & {\rm otherwise}
    \end{cases}
\end{equation} 
and collect them into an $r \times M$ matrix $S$:
\begin{equation}
    S\equiv\left(\vec{p}_1,\,\vec{p}_2,\,\dots,\,\vec{p}_r\right)^\intercal
\label{defS}
\end{equation}
We then define the $r$-dimensional Kronecker delta over $\mathbb{F}_2$ as
\begin{equation}
    \delta^{(r)}(S\vec{x})
    \equiv\prod_{s=1}^r\delta(\vec{p}_s^{\,\intercal} \vec{x})
    =\prod_{s=1}^r  \delta_{\,\sum_{i \in C_s} (\vec{x})_i,\,{\rm even}}
\end{equation}
where the second equality follows from~(\ref{defp}). With this notation, (\ref{eq:trpiG_result}) becomes:
\begin{equation}
    \bra{G}U(\pi)\ket{G}
    =\frac{2^r}{2^M}\sum_{\vec{x}\,\in\, \mathbb{F}_2^M}
       (-)^{\phi(\vec{x})\,+\,\sum_{s=1}^r \sum_{(i>j)\in C_s}(\vec{x})_i (A\vec{x})_j} \times
        \delta^{(r)}(S\vec{x})\,\, \delta^{(r)}(SA\vec{x})
    \label{eq:trpiG_result2}
\end{equation}
For a final simplification, observe that $\ker{S}={\rm im}(1+P)$, where $P$ is the action of $\pi$ on $\mathbb{F}_2^M$ defined in (\ref{defP}). One way to verify this identification is by observing that, manifestly, ${\rm im}(1+P)\subseteq \ker{S}$ and $\dim {\rm im}(1+P)= M-r = \dim \ker{S}$. Consequently, the factor $\delta^{(r)}(S\vec{x})$ restricts the summation to ${\rm im}(1+P)$:
\begin{equation}
    \bra{G}U(\pi)\ket{G}
    =\frac{2^r}{2^M}\sum_{\vec{x}\,\in\, {\rm im}(1+P)}
       (-)^{\phi(\vec{x})\,+\,\sum_{s=1}^r \sum_{(i>j)\in C_s}(\vec{x})_i (A\vec{x})_j}
        \times \delta^{(r)}(SA\vec{x})
    \label{eq:trpiG_resultFINAL}
\end{equation}

At a first glance, formulae~(\ref{eq:trpiG_result}) and (\ref{eq:trpiG_resultFINAL}) may not seem illuminating. Yet they turn out to be very practical in computations of multi-invariants. We illustrate the use of equation~(\ref{eq:trpiG_resultFINAL}) with an explicit example in Section~\ref{sec:explicit}, as well as several multi-invariant calculations in Appendix~\ref{appendix:MI}. 

\subsection{Comparison with the general approach}
\label{subsec:comparetwoapproaches}
Formula~(\ref{eq:trpiG_resultFINAL}) appears very different from equation~(\ref{oldapproach}), which is the product of Sections~\ref{sec: Innerprod} and \ref{sec: Arf_inv}. We presently verify that the expressions are in fact equal and, in the process, clarify the conceptual distinction between the two approaches. 

As a first step, we rewrite (\ref{oldapproach}) in the form:
\begin{equation}
    \bra{G}U(\pi)\ket{G}=\frac{1}{2^M}\sum_{\vec{x}\,\in\,\mathbb{F}_2^M}(-)^{\phi(\vec{x})+\phi(P\vec{x})}
    \label{smallrewrite}
\end{equation}
To make contact with~(\ref{eq:trpiG_resultFINAL}), we organize the summation according to the decomposition
\begin{equation}
\mathbb{F}_2^M 
\cong \frac{\mathbb{F}_2^M}{\ker(1+P)} \oplus \ker(1+P)\,,
\label{f2mdecomp}
\end{equation}
where the quotient is isomorphic to ${\rm im}(1+P)$. Explicitly, for each coset in the quotient we pick a representative $\vec{x}_y$, which satisfies $(1+P) \vec{x}_y = \vec{y} \in {\rm im}(1+P)$. Meanwhile, $\ker (1+P)$ is spanned by the vectors $\vec{p}_s$, which were defined in equation~(\ref{defp}). Thus, to organize a sum according to (\ref{f2mdecomp}) is to replace:
\begin{equation}
    \sum_{\vec{x}\,\in\, \mathbb{F}_2^M}f(\vec{x})
    \quad \longrightarrow \quad 
    \sum_{\vec{y}\,\in\,{\rm im}(1+P)}\,\sum_{\vec{t}\,\in\,\mathbb{F}_2^{\!\phantom{|}r}}
    f\Big(\vec{x}_y + \sum_{s=1}^r t_s\, \vec{p}_s\Big)
    \label{sumrewrite}
\end{equation}
Applying this to~(\ref{smallrewrite}) yields:
\begin{align}
\label{eq: LHS_rendered}
    \frac{1}{2^M}\sum_{\vec{x}\,\in\, \mathbb{F}_2^M} 
    (-)^{\phi(\vec{x})+\phi(P\vec{x})} 
    &=\frac{1}{2^M}\sum_{\vec{x}\,\in\, \mathbb{F}_2^M} (-)^{\phi(\vec{x}+ P\vec{x})+\vec{x}^\intercal A (\vec{x}+P\vec{x})} \notag \\
    &=\frac{1}{2^M} \sum_{\vec{y}\,\in\,{\rm im}(1+P)}\sum_{\vec{t}\,\in\,\mathbb{F}_2^{\!\phantom{|}r}}
    (-)^{\phi(\vec{y})\,+\,(\vec{x}_y+\sum_{s=1}^r t_s \vec{p}_s)^\intercal A \vec{y}}\notag \\
    &=\frac{2^r}{2^M} \sum_{\vec{y}\,\in\,{\rm im}(1+P)}
    (-)^{\phi(\vec{y})\,+\,\vec{x}_y^\intercal A \vec{y}} 
    \left(
    \frac{1}{2^r}\sum_{\vec{t}\,\in\,\mathbb{F}_2^{\!\phantom{|}r}}(-)^{\vec{t}^\intercal S A \vec{y}}
    \right)\notag \\
    &=\frac{2^r}{2^M} \sum_{\vec{y}\,\in\,{\rm im}(1+P)}
    (-)^{\phi(\vec{y})\,+\,\vec{x}_y^\intercal A \vec{y}} 
    \times \delta^{(r)}(SA\vec{y})
\end{align}
The first line uses the key property of $\phi(\vec{x})$ (equation~(\ref{cohomology})) and the fact that $\vec{x}^\intercal A \vec{x}=0$ because $A$ is symmetric; the second line implements (\ref{sumrewrite}); the third line uses definition~(\ref{defS}); the fourth line recognizes the paranthesis as the discrete Fourier transform of the delta function:
\begin{equation}
     \frac{1}{2^r}\sum_{\vec{t}\,\in\,\mathbb{F}_2^{\!\phantom{|}r}}(-)^{\vec{t}^\intercal \vec{v}}
    =\delta^{(r)}(\vec{v})   
    =\begin{cases}
    1\quad {\rm if}~\vec{v}=\vec{0} \\
    0\quad \text{otherwise}
    \end{cases}\qquad 
    {\rm for}~{\vec{v}\,\in\,\mathbb{F}_2^{\!\phantom{|}r}}
    \label{ddelta}
\end{equation}

Equation~\eqref{eq: LHS_rendered} reproduces \eqref{eq:trpiG_resultFINAL} if for all $\vec{y}\in \ker(S A)$
\begin{equation}
\label{eq: microscopic match}
    \sum_{s=1}^{r} \sum_{(i>j)\in C_s}(\vec{y})_i (A\vec{y})_j
    = \vec{x}_y^\intercal A \vec{y}\,,
\end{equation}
which in turn follows if for every cycle $C_s$:
\begin{equation}
\label{eq:micromicro}
    \sum_{(i>j)\in C_s} (\vec{y})_i (A\vec{y})_j
    = \sum_{j \in C_s} (\vec{x}_y)_j (A \vec{y})_j
\end{equation}
To verify~(\ref{eq:micromicro}) in a transparent manner, fix the labeling of the qubits so that $C_s = (1\, 2\,\dots\, l)$. (Recall that equation~(\ref{eq:trpiG_resultFINAL}) only assumes that the qubits are indexed such that each cycle permutes them `in order'; we might as well index the qubits in $C_s$ with $1 \leq i \leq l$.)  In components, definition $(1+P)\vec{x}_y=\vec{y}$ says that for $2 \leq i \leq l$, $(\vec{y})_i=(\vec{x}_y)_i+(\vec{x}_y)_{i-1}$. We therefore have:
\begin{align}
    \sum_{(i>j)\in C_s} (\vec{y})_i (A\vec{y})_j 
    & \, = \sum_{(i>j)\in C_s}
    \Big((\vec{x}_y)_{i}+(\vec{x}_y)_{i-1}\Big)(A\vec{y})_j
    = \sum_{j \in C_s}
    \Big((\vec{x}_y)_l+(\vec{x}_y)_j\Big)(A\vec{y})_j 
    \notag \\
    & \,= \sum_{j\in C_s} (\vec{x}_y)_j (A\vec{y})_j
\end{align}
In the final line, we dropped a term proportional to $\sum_{j\in C_s}(A\vec{y})_j=\vec{p}_s^{\,\intercal} A \vec{y}=(SA\vec{y})_s$ because $\vec{y}\in \ker(SA)$.

This completes the proof that the approaches in Sections~\ref{sec: Innerprod} and \ref{sec: Arf_inv} and in Section~\ref{subsec:tracecycles} yield the same answer.

\paragraph{Summary of comparison}
We have given two ways to compute:
\begin{equation}
    \bra{G}U(\pi)\ket{G}
    =\frac{1}{2^M}\sum_{\vec{x}\,\in\,\mathbb{F}_2^M}(-1)^{\vec{x}^\intercal (A+P^\intercal\! A P)^{\suppertr} \vec{x}}
\end{equation}
When following Sections~\ref{sec: Innerprod} and \ref{sec: Arf_inv}, we organize the sum over $\vec{x}\in\mathbb{F}_2^M$ according to the isomorphism:
\begin{equation}
\mathbb{F}_2^M 
\cong \frac{\mathbb{F}_2^M}{\ker(A+P^\intercal A P)} \oplus \ker(A+P^\intercal A P)
\end{equation}
This is evident, for example, in equation~(\ref{allfactors}) with the substitution $A \to A+P^\intercal A P$. 

In contrast, Section~\ref{subsec:tracecycles} organizes the same sum by exploiting the isomorphism
\begin{equation}
\mathbb{F}_2^M 
\cong \frac{\mathbb{F}_2^M}{\ker(1+P)} \oplus \ker(1+P),\,
\tag{\ref{f2mdecomp}}
\end{equation}
which is manifest in equation~(\ref{eq:trpiG_resultFINAL}). 

\subsection{Explicit example: Complete graph $\mathcal{K}_4$}
\label{sec:explicit}
Let $\mathcal{K}_4$ denote the complete graph on $M=4$ vertices. We inspect the equation:
\begin{equation}
    \bra{+}^{\otimes 4}\ket{\mathcal{K}_4} = (-)^\Delta \times 2^{-({\rm rank}A)/2} 
    = -\frac{1}{4} \equiv \bra{G}U(\pi)\ket{G}
    \label{whatweinspect}
\end{equation}
The first equalities are taken from Section~\ref{sec: Arf_inv}; quantities $\Delta = 1$ and ${\rm rank}\,A = 4$ for $G = \mathcal{K}_4$ were computed in the last part of Section~\ref{sec: arf_ex}. 

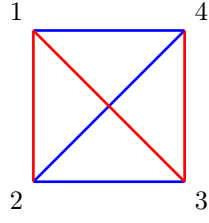
\begin{figure}[t]
    \centering
    \begin{tikzpicture}[scale=2]
    
        \coordinate (A) at (0,0);
        \coordinate (B) at (1,0);
        \coordinate (C) at (1,1);
        \coordinate (D) at (0,1);
        \node[below left]  at (A) {\small 2};
        \node[below right] at (B) {\small 3};
        \node[above right] at (C) {\small 4};
        \node[above left]  at (D) {\small 1};

        \draw[line width=1pt, blue] (A) -- (B); 
        \draw[line width=1pt, blue] (A) -- (C); 
        \draw[line width=1pt, blue] (D) -- (C); 

        \draw[line width=1pt, red] (A) -- (D);
        \draw[line width=1pt, red] (B) -- (D);
        \draw[line width=1pt, red] (B) -- (C);
    \end{tikzpicture}

    \caption{The complete graph $\mathcal{K}_4$ as a union of graph $G$ (red) and $\pi (G)$ (blue), where $\pi=(1\, 2\, 3\, 4)$ permutes the vertices of $G$. We have drawn the figure such that $\pi$ can be visualized as an anti-clockwise rotation by $90^\circ$.}
    \label{fig: ex K_4}
\end{figure}

The last equality in (\ref{whatweinspect}) \emph{defines} a new graph $G$ whose union with $\pi (G)$ reproduces $\mathcal{K}_4$. Here $\pi(G)$ is obtained from $G$ by permuting its vertices with $\pi = (1\, 2\, 3\, 4)$. The right hand side of (\ref{whatweinspect}) is in a form suitable to the approach from Section~\ref{subsec:tracecycles}. We presently calculate 
\begin{equation}
    \bra{G}U(\pi)\ket{G}=\frac{1}{2^M}\sum_{\vec{x}\,\in\, \mathbb{F}_2^M} (-1)^{\phi(\vec{x})+\phi(P\vec{x})}
\end{equation}
by following that formalism. Our setup is illustrated in Figure~\ref{fig: ex K_4}. 

\paragraph{The calculation}
Our starting point is
\begin{equation}
\label{eq: LHS_rendered2}
    \frac{1}{2^M}\sum_{\vec{x}\,\in\, \mathbb{F}_2^M} 
    (-)^{\phi(\vec{x})+\phi(P\vec{x})} 
    =\frac{1}{2^M} \sum_{\vec{y}\,\in\,{\rm im}(1+P)}
    2^r \times (-)^{\phi(\vec{y})\,+\,\vec{x}_y^\intercal A \vec{y}} 
    \times \delta^{(r)}(SA\vec{y})\,,
\end{equation}
which is equation~(\ref{eq: LHS_rendered}) with the substitution~(\ref{eq: microscopic match}). The parameters are $M=4$ and $r=1$; the adjacency matrix $A$, the permutation matrix $P=P(\pi)$, and the object $S$ defined in equation~(\ref{defS}) are: 
\begin{equation}
    A=\begin{bmatrix}
        0 & 1 & 1 & 0 \\ 
        1 & 0 & 0 & 0 \\
        1 & 0 & 0 & 1 \\ 
        0 & 0 & 1 & 0
    \end{bmatrix}
    \qquad 
    P=\begin{bmatrix}
        0 & 0 & 0 & 1 \\ 
        1 & 0 & 0 & 0 \\
        0 & 1 & 0 & 0 \\ 
        0 & 0 & 1 & 0
    \end{bmatrix}
    \qquad
    S=\begin{bmatrix}
        1 & 1 & 1 & 1
    \end{bmatrix}
\end{equation} 

On the right hand side of~(\ref{eq: LHS_rendered2}), each $\vec{y}$ selects a stabilizer $O_{\vec{y}} = K_1^{y_1} K_2^{y_2} K_3^{y_3} K_4^{y_4}$ of $\ket{G}$. Condition~$\vec{y}\in {\rm im}(1+P)$ guarantees that
\begin{equation}
    O_{\vec{y}}=(-)^{\phi(\vec{y})}\cdot\left(X_1^{y_1} Z_1^{(A\vec{y})_1}\right)\otimes \left(X_2^{y_2} Z_2^{(A\vec{y})_2}\right)\otimes \left(X_3^{y_3} Z_3^{(A\vec{y})_3}\right) \otimes \left(X_4^{y_4} Z_4^{(A\vec{y})_4}\right)
\end{equation}
have an even number of $X$'s in the one cycle $(1\, 2\, 3\, 4)$ of $\pi$. The factor $\delta^{(1)}(SA\vec{y})$ ensures that $O_{\vec{y}}$ have an even number of $Z$'s in the cycle. 
For each $\vec{y}$ that satisfies both these conditions, the magnitude of the contribution is $2^r = 2$ because $\vec{y}$ represents the coset $\vec{x}_y+\ker(1+P) = \{\vec{x}_y,\vec{x}_y+(1,1,1,1)^\intercal\}$ with 2 elements.
The sign of the contribution of $\vec{y}$ is determined by the combination $\phi(\vec{y}) + \vec{x}_y^\intercal A \vec{y}$. 

Table~\ref{tab:grouping_wrt_perminvspace} groups the summands $\vec{x}$ on the left hand side of (\ref{eq: LHS_rendered2}) according to their role in the right hand side summation. The lower half of the table gathers those cosets $\vec{x}_y+\ker(1+P)$, which fail the condition imposed by $\delta^{(1)}(SA\vec{y})$; such cosets make zero net contribution to the summation.

\begin{table}[htbp]
    \centering
    \caption{Explicit terms in the summations on both sides of (\ref{eq: LHS_rendered2}), applied to the setup in Figure~\ref{fig: ex K_4}.}
    \label{tab:grouping_wrt_perminvspace}

    \begin{tabular}{|c|cccc||cccc|c|}
        \hline
        \multirow{2}{*}{$(-)^{\phi(\vec{x})+\phi(P\vec{x})}$} 
        & \multicolumn{4}{c||}{$\vec{x}$} 
        & \multicolumn{4}{c|}{$\vec{y} = (1+P) \vec{x}$} 
        & \multirow{2}{*}{$2^r \times (-)^{\phi(\vec{y})\,+\,\vec{x}_y^\intercal A \vec{y}} 
    \times \delta^{(r)}(SA\vec{y})$} \\
        \cline{2-9}
        & $x_1$ & $x_2$ & $x_3$ & $x_4$ & $y_1$ & $y_2$ & $y_3$ & $y_4$ & \\ 
        \hline\hline

        $+1$ & 0 & 0 & 0 & 0 & \multirow{2}{*}{0} & \multirow{2}{*}{0} & \multirow{2}{*}{0} & \multirow{2}{*}{0} & \multirow{2}{*}{$+2$} \\
        \cline{1-5}
        $+1$ & 1 & 1 & 1 & 1 &  &  &  &  &  \\ \hline
        
        $-1$ & 1 & 0 & 1 & 0 & \multirow{2}{*}{1} & \multirow{2}{*}{1} & \multirow{2}{*}{1} & \multirow{2}{*}{1} & \multirow{2}{*}{$-2$} \\
        \cline{1-5}
        $-1$ & 0 & 1 & 0 & 1 &  &  &  &  &  \\ \hline
        
        $-1$ & 1 & 1 & 0 & 0 & \multirow{2}{*}{1} & \multirow{2}{*}{0} & \multirow{2}{*}{1} & \multirow{2}{*}{0} & \multirow{2}{*}{$-2$} \\
        \cline{1-5}
        $-1$ & 0 & 0 & 1 & 1 &  &  &  &  &  \\ \hline
        
        $-1$ & 1 & 0 & 0 & 1 & \multirow{2}{*}{0} & \multirow{2}{*}{1} & \multirow{2}{*}{0} & \multirow{2}{*}{1} & \multirow{2}{*}{$-2$} \\
        \cline{1-5}
        $-1$ & 0 & 1 & 1 & 0 &  &  &  &  &  \\ \hline\hline
        
        $-1$ & 0 & 1 & 1 & 1 & \multirow{2}{*}{1} & \multirow{2}{*}{1} & \multirow{2}{*}{0} & \multirow{2}{*}{0} & \multirow{2}{*}{0} \\
        \cline{1-5}
        $+1$ & 1 & 0 & 0 & 0 &  &  &  &  &  \\ \hline
        
        $-1$ & 1 & 1 & 1 & 0 & \multirow{2}{*}{1} & \multirow{2}{*}{0} & \multirow{2}{*}{0} & \multirow{2}{*}{1} & \multirow{2}{*}{0} \\
        \cline{1-5}
        $+1$ & 0 & 0 & 0 & 1 &  &  &  &  &  \\ \hline
        
        $-1$ & 1 & 0 & 1 & 1 & \multirow{2}{*}{0}  & \multirow{2}{*}{1} & \multirow{2}{*}{1} & \multirow{2}{*}{0} & \multirow{2}{*}{0} \\
        \cline{1-5}
        $+1$ & 0 & 1 & 0 & 0 &  &  &  &  &  \\ \hline
        
        $-1$ & 1 & 1 & 0 & 1 & \multirow{2}{*}{0} & \multirow{2}{*}{0} & \multirow{2}{*}{1} & \multirow{2}{*}{1} & \multirow{2}{*}{0} \\
        \cline{1-5}
        $+1$ & 0 & 0 & 1 & 0 &  &  &  &  &  \\ \hline
    \end{tabular}
\end{table}

\paragraph{Remark}
Section~\ref{subsec:tracecycles} studies the expression:
\begin{equation}
     \frac{1}{2^M}\sum_{\vec{y}\,\in\, \mathbb{F}_2^M} {\rm tr}\left(U(\pi)O_{\vec{y}}\right)
     =  \bra{G}U(\pi)\ket{G}
     = \sum_{\vec{x}\,\in\, \mathbb{F}_2^M} \braket{G|\vec{x}} \bra{\vec{x}}U(\pi)\ket{G}
     \label{macroscopic}
\end{equation}
In the left equality we used~(\ref{allstabs}) and on the right we inserted the resolution of the identity in terms of $Z$-eigenstates. The manipulations in Section~\ref{subsec:comparetwoapproaches} allow us to decompose~(\ref{macroscopic}) into individual summands. We find that
\begin{equation}
    \frac{1}{2^M}\, {\rm tr}\left(U(\pi)O_{\vec{y}}\right)
    \,= \sum_{\vec{x}\,\in\, \vec{x}_y+\,\ker(1+P)} \bra{\vec{x}}U(\pi)\ket{G}\braket{G|\vec{x}}
    \label{microresult}
\end{equation}
if $\vec{y}\in {\rm im}(1+P)$ and 0 otherwise. 

To see this, repeat the steps from equation~(\ref{trpiG_expanded}) to (\ref{eq:trpiG_result2}) at the level of an individual summand and substitute~(\ref{eq: microscopic match}) in the final expression:
\begin{equation}
    \frac{1}{2^M}\, {\rm tr}\left(U(\pi)O_{\vec{y}}\right)
    = 
    \frac{2^r}{2^M}\, (-)^{\phi(\vec{y})\,+\,\vec{x}_y^\intercal A \vec{y}}
    \times \delta^{(r)}(S\vec{y})\,\, \delta^{(r)}(SA\vec{y})
\end{equation}
Since the $\delta^{(r)}(S\vec{y})$ requires $\vec{y}\in {\rm im}(1+P)$ for a nonvanishing result, it suffices to prove
\begin{equation}
   Q_L \equiv 
    \frac{2^r}{2^M}\, (-)^{\phi(\vec{y})\,+\,\vec{x}_y^\intercal A \vec{y}}
    \times \delta^{(r)}(SA\vec{y})
    = \sum_{\vec{x}\,\in\, \vec{x}_y+\,\ker(1+P)} \bra{\vec{x}}U(\pi)\ket{G}\braket{G|\vec{x}} 
    \equiv Q_R
\end{equation}
assuming $\vec{y}\in {\rm im}(1+P)$. This holds because:
\begin{align}
Q_L 
  &\, = 
    \frac{2^r}{2^M} \,
    (-)^{\phi(\vec{y})\,+\,\vec{x}_y^\intercal A \vec{y}} 
    \left(
    \frac{1}{2^r}\sum_{\vec{t}\,\in\,\mathbb{F}_2^{\!\phantom{|}r}}  (-)^{\vec{t}^\intercal S A \vec{y}}
    \right)
  = 
   \frac{1}{2^M} \sum_{\vec{t}\,\in\,\mathbb{F}_2^{\!\phantom{|}r}}
   (-)^{\phi(\vec{y})+(\vec{x}_y^\intercal + \vec{t}^\intercal S) A \vec{y}} 
   \\
 &\, = 
   \frac{1}{2^M}\sum_{\vec{x}\,\in\, \vec{x}_y+\,\ker(1+P)}(-)^{\phi((1+P)\vec{x})+\vec{x}^\intercal A (1+P) \vec{x}}
   = 
   \frac{1}{2^M}\sum_{\vec{x}\,\in\, \vec{x}_y+\,\ker(1+P)}(-)^{\phi(\vec{x}+P\vec{x})+\vec{x}^\intercal A P \vec{x}}
   \nonumber \\
 &\, =
   \frac{1}{2^M}\sum_{\vec{x}\,\in\, \vec{x}_y+\,\ker(1+P)}(-)^{\phi(\vec{x})+\phi(P\vec{x})}
   = 
   \sum_{\vec{x}\,\in\, \vec{x}_y+\,\ker(1+P)}\frac{(-)^{\phi(P\vec{x})}}{2^{M/2}}\cdot\frac{(-)^{\phi(\vec{x})}}{2^{M/2}}
   = Q_R
   \nonumber
\end{align}
The first line uses the discrete Fourier transform of the delta function~(\ref{ddelta}). The second line uses the defining property $(1+P)\vec{x}_y = \vec{y}$ and recognizes $S^\intercal \vec{t}$ with $\vec{t} \in \mathbb{F}_2^{\,r}$ as a parameterization of $\ker (1+P)$, then exploits $\vec{x}^\intercal A \vec{x} = 0$. The third line uses~(\ref{cohomology}) as well as explicit expressions for $\bra{\vec{x}}U(\pi)\ket{G}$ and $\braket{G|\vec{x}}$. This proves equation~(\ref{microresult}). 

\section{Discussion} 
\label{sec:discuss}
Our study of the inner product of two graph states took us on a rather picaresque journey. In what follows we summarize the main findings one more time, now equipped with the proper vocabulary. We then outline potential applications. 

\paragraph{Summary of results} The inner product $\braket{G' | G''}$ can be expressed as the partial amplitude $\bra{+}^{\otimes M}\ket{G}$, where graph $G$ is the \emph{symmetric difference} of $G'$ and $G''$. Quantity~$\bra{+}^{\otimes M}\ket{G}$ has the following characteristics:
\begin{itemize}
\item The amplitude $\bra{+}^{\otimes M}\ket{G}$ vanishes if and only if graph $G$ contains an induced subgraph $H$ with an odd number of edges such that every vertex of $G$ has an even number of neighbors in $H$.
\item When $|\bra{+}^{\otimes M}\ket{G}|^2 \neq 0$ it equals $2^{-{\rm rank} A}$, where $A$ is the adjacency matrix of graph $G$ and the rank is understood over the binary field $\mathbb{F}_2$.
\item The $\pm$ phase in $\bra{+}^{\otimes M}\ket{G} = \pm 2^{-({\rm rank} A)/2}$ is determined by the Arf invariant of the quadratic refinement $\phi$ (equation~\ref{phiexplicit1}) of $A$, understood as a bilinear form. 
\item When a general linear transformation brings the adjacency matrix $A$ to its canonical, symplectic form, the Arf invariant counts (mod 2) symplectic blocks on which $\phi$ has a distinct behavior: case~($\sharp$) below equation~(\ref{allfactors}).
\item The same general linear transformation motivates the introduction of nonlocal effective qubits, in terms of which every graph state $\ket{G}$ is a product of Bell pairs and unentangled ancillae. The effective qubits define an alternative tensor factorization of the Hilbert space, which is distinct from the tensor factors set by the local physical constituents of $\ket{G}$. 
\item Section~\ref{sec:compute_permut} complements the above results with an additional formalism, which is specifically designed for computing expectation values of qubit-wise permutations in graph states, $\bra{G}U(\pi)\ket{G}$. This formalism is useful for computing multi-invariants.
\end{itemize}

\subsection{Potential applications}

\paragraph{Tentative interpretation in Measurement-Based Quantum Computation} MBQC \cite{PhysRevLett.86.5188} is a general formalism for conducting quantum computations. In contrast to the circuit model, in which a quantum state undergoes transformations effected by gates, MBQC realizes quantum computations by subjecting a cleverly chosen resource state to adaptive single-qubit measurements. The computation is determined by the choice of measurement bases whereas the computational output is an appropriate combination of the qubit-wise measurement outcomes. 

The results of this paper are likely to inform Measurement-Based Quantum Computation because graph states are prototypical MBQC resource states \cite{PhysRevA.68.022312}. In routine presentations of MBQC, one begins with a rigid resource graph state---most often one whose underlying graph is a regular lattice in two dimensions (sufficient for universal quantum computation \cite{Nest:2006rge}) or in three dimensions (sufficient for fault tolerance \cite{Raussendorf:2005suv}). But the subsequent step of the computation strips off ``unnecessary'' qubits using $Z$-basis measurements to reveal a far smaller remaining graph state; call it $\ket{G}$. Because the properties of $G$ reflect the structure of the intended computation, it is intriguing to apply our results to this graph.

To fully flesh out what our results say about MBQC merits a separate research project, which two of us have already initiated \cite{withyichen}. Here we limit ourselves to these preliminary observations:
\begin{enumerate}
\item In the canonical MBQC equivalent of the trivial circuit computation (MBQC emulation of the identity), $G$ is an odd-membered chain or an even-membered loop\footnote{The realization on even-membered loops achieves ``super-dense coding'': two classical bits of computational output for each virtual qubit employed in the computation \cite{Wong_2024}.} or a disjoint union thereof and all measurements are done in the $X$-basis. In this instance, the ``computational output'' is---indeed, must be---deterministic: it is the sum of single-qubit measurements at alternating locations. A lookup in Section~\ref{sec: arf_ex} reveals that this computational output consists of the special stabilizers~(\ref{kernelstab}), which are associated with the kernel of the adjacency matrix and which in Section~\ref{sec: arf_interpret} gave rise to unentangled effective qubits. 
\label{point1}
\item When a graph state of more general architecture is measured in the $X$-basis, stabilizers~(\ref{kernelstab}) likewise constitute deterministic computational outputs. Under mild additional assumptions, this setup is the MBQC realization of classical computations built up from XOR (``classical CNOT'') gates.
\item Reference~\cite{Wong_2024} interpreted the deterministic computational outputs from point~\ref{point1} above as fluxes of a background $\mathbb{Z}_2 \times \mathbb{Z}_2$ gauge field. In that language, truly quantum (non-classical) computations in the MBQC formalism involve measuring the flux in a superposition basis. In the formalism developed in Section~\ref{sec: arf_interpret}, this means measuring the unentangled effective qubits depicted in Figure~\ref{appfig:new_graph} in a general basis, which depends on the intended MBQC computation.
\label{point3}
\item The local measurements in MBQC are adaptive, meaning that later measurement bases depend on prior measurement results. Based on our preliminary examination, the pattern of adaptations can be encoded in appropriately chosen hyperbolic blocks $\left\{\vec{x}_i, \vec{y}_i\right\}$ of the symplectic basis~(\ref{directsumdecomposition}-\ref{appeqn: A_new}).
\end{enumerate}
A comprehensive investigation of this subject is certainly outside the scope of this paper. We hope that these quick observations justify our belief that the results of this paper are relevant and helpful for understanding MBQC. 

One promising application is to the issue of the quantum advantage: What makes quantum computers more powerful than classical computers? Applying our results to MBQC suggests a tentative answer: \emph{nonlocality}. Indeed, in a formalism in which computational inputs are \emph{locally} inserted and the computation proceeds by \emph{local} measurements, the computational outputs reside in \emph{nonlocal} effective qubits. The latter are defined by an alternative tensor factorization of the Hilbert space, which bears no simple relation to the tensor factors set by the physical, localized qubits. As a thought exercise, one can imagine redefining the notion of locality such that the effective qubits of Figure~\ref{appfig:new_graph} become local. In this hypothetical viewpoint, the computational outputs of MBQC are read off by local measurements, but then the computational inputs---locally specified in the conventional picture---are encoded by highly nonlocal correlations among the effective qubits. In short, MBQC allows one to localize computational inputs \emph{or} computational outputs, but not both.

We aim to develop these thoughts more fully in upcoming publications.

\paragraph{Algebraic topology} Point~\ref{point3} in the list of MBQC observations suggests that the involvement of the Arf invariant in our analysis may not be a mere accident of binary field algebra. In algebraic topology, the Arf invariant partly characterizes spin structures on two-dimensional manifolds. For physicists, this means setting periodic (Ramond, R) or antiperiodic (Neveu-Schwarz, NS) boundary conditions around closed loops for fermions living on the given manifold. One can consider virtual fermions as test particles whose transport detects fluxes of a $\mathbb{Z}_2$ gauge theory: trivial flux for the periodic (R) and non-trivial flux for the antiperiodic (NS) boundary conditions. In the context of an MBQC computation, a pair of such flux states in a quantum superposition naturally encodes an arbitrary $SU(2)$ transformation applied to a virtual qubit:
\begin{align}
\label{utophi}
U & = c_{00} I + i c_{01} X + i c_{10} Z + i c_{11} Y 
\\
\longleftrightarrow \quad
\ket{\Phi} & = 
c_{00} \ket{\textrm{R, R}} 
      + i c_{01} \ket{\textrm{R, NS}} + i c_{10} \ket{\textrm{NS, R}} + i c_{11} \ket{\textrm{NS, NS}}
\nonumber
\end{align}
(The coefficients are real and satisfy $c_{00}^2 + c_{01}^2 + c_{10}^2 + c_{11}^2 = 1$.) Relationship~(\ref{utophi}) was used in \cite{Wong_2024} to reformulate MBQC on one-loop resource states as a measurement of $\mathbb{Z}_2 \times \mathbb{Z}_2$ flux in a superposition basis.

To recap, a single-qubit quantum computation can be faithfully represented as a quantum superposition of spin structures on a manifold with two independent, nonintersecting and noncontractible loops. In its unexpected appearance in Section~\ref{sec: Arf_inv}, does the Arf invariant hint that multi-qubit quantum computations can likewise be geometrized into spin structures---only on more complicated manifolds? 
If we tentatively adopt this point of view, the following ideas suggest themselves:
\begin{itemize}
\item The adjacency matrix of graph $G$ plays the role of the intersection form on the first homology $H_1(\mathcal{M}, \mathbb{Z}_2)$ of an auxiliary manifold $\mathcal{M}$. The calculation we did in Section~\ref{sec:arf} brought the intersection matrix of $\mathcal{M}$ to its canonical, symplectic form \cite{farbmargalit}. 
\item The nonlocal effective qubits, which we defined in Section~\ref{sec: arf_interpret} and depicted in Figure~\ref{appfig:new_graph}, determine the topology of $\mathcal{M}$: 
\begin{itemize}
\item The number of punctures is one plus the number of unentangled effective qubits $\underline{\ket{\pm}}$. The extra puncture is due to $\mathbb{Z}_2$-flux Gauss law: the loop, which bounds a disk containing all the punctures cannot support nontrivial flux because it is contractible.
\item The number of handles (the genus) equals the number of effective Bell pairs.
\end{itemize}
\item Handles with periodic-periodic (R-R) boundary conditions are singlets under $\mathbb{Z}_2$-modular transformations, so they correspond to case~($\sharp$) below equation~(\ref{allfactors}). Handles with the other boundary conditions (NS-NS, NS-R, R-NS) form triplets under $\mathbb{Z}_2$-modular transformations and correspond to option ($\flat$).
\item The Arf invariant counts handles with R-R boundary conditions. 
\item The Arf invariant is not defined when the flux around any puncture $\gamma$ is nontrivial because a cycle from an R-R type pair can be combined with $\gamma$ and converted into an NS-R type pair. This rationale mirrors the logic in the final paragraphs of Section~\ref{sec: arf_interpret}. 
\end{itemize}
The above ideas are still tentative; we aim to make them precise in upcoming publications. A potentially useful set of tools toward that goal, based on topological quantum field theory, was developed in Reference~\cite{Wong:2023bhs}.  

\paragraph{A testbed for multi-invariants}
The formalism outlined in Section~\ref{sec:compute_permut} affords a far greater breadth of multi-invariant calculations than were previously feasible. One example is this unpublished result, which was proven using the technique of Section~\ref{sec:compute_permut} \cite{treegraphME}: The most fine-grained (one qubit per party) index-$n$ R{\'e}nyi multi-entropy of an $M$-qubit tree graph state is a polynomial in $(1/n)$ of degree $(M-2)$ or lower. We illustrate and apply this result in Appendix~\ref{sec: C4}. Looking ahead, we hope that the extra computational flexibility afforded by this group-theoretic approach paves the way to a better understanding of multi-invariants.

\acknowledgments
We thank Jonathan Harper, Xin-Xiang Ju, Dimitrios Patramanis, Xiaoliang Qi, Sirui Shuai and Zhenbin Yang for discussions and Sriram Akella, Abhijit Gadde and Jay Pandey for sharing a draft of their paper~\cite{akella2026multiinvariantsstabilizerstates} prior to arXiv announcement. 
BC thanks the organizers of the workshops ``Discussions on Quantum Spacetime'' held at LMSI (Mumbai, India) and ``AI for Physics'' held at the Aspen Center for Physics while XLW thanks the organizers of the 20$^{\rm th}$ Asian Winter School on Strings, Particles and Cosmology held at IISER Bhopal (India), where part of this work was completed. This research was supported by an NSFC grant number 12042505. 

\appendix

\section{Wavefunction of graph states in the $X$-basis}
\label{app:wfx}
The method presented in Section~\ref{sec: Arf_inv} allows us to efficiently compute the full 
wavefunction of $\ket{G}$ in the $X$-basis. We denote $X$-eigenstates with 
\begin{equation}
    \ket{(-)^{\vec{u}}} \equiv \ket{(-)^{u_1}}\otimes\ket{(-)^{u_2}}\otimes\cdots\otimes\ket{(-)^{u_M}},
\end{equation}
where $\vec{u}$ is a length-$M$ binary bitstring and 
\begin{equation}
    \ket{(-)^0}=\ket{+}=\frac{\ket{0}+\ket{1}}{\sqrt{2}},\quad \ket{(-)^1}=\ket{-}=\frac{\ket{0}-\ket{1}}{\sqrt{2}}.
\end{equation}
The goal is to compute $\braket{(-)^{\vec{u}}|G}$.

Observe that the $X$-eigenstate $\ket{(-)^{\vec{u}}}$ is related to the $Z$-eigenstates $\ket{\vec{x}}$ via:
\begin{equation}
    \ket{(-)^{\vec{u}}}=
    \frac{1}{2^{M/2}}\sum_{\vec{x}\,\in\, \mathbb{F}_2^M} (-)^{\vec{u}\cdot\vec{x}}\ket{\vec{x}}
\end{equation}
Using \eqref{czactiononx}, the wavefunction of $\ket{G}$ in the $Z$-eigenbasis is:
\begin{equation}
    \ket{G}=
    \frac{1}{2^{M/2}}\sum_{\vec{x}\,\in\, \mathbb{F}_2^M} (-)^{\phi(\vec{x})}\ket{\vec{x}},
\end{equation}
Quantities $\phi(\vec{x})=\vec{x}^\intercal A^{\uppertr} \vec{x}$ and $A^{\uppertr}$ were defined in the main text in equations~(\ref{phiexplicit1}) and (\ref{defa}). Therefore, the amplitude $\braket{(-)^{\vec{u}}|G}$ we are seeking is:
\begin{equation}
\label{amplitudephiu}
    \braket{(-)^{\vec{u}}|G}
    =\frac{1}{2^M}\sum_{\vec{x}\,\in\, \mathbb{F}_2^M} (-)^{\phi(\vec{x})+\vec{u}\cdot\vec{x}}
    \equiv\frac{1}{2^M}\sum_{\vec{x}\,\in\, \mathbb{F}_2^M}(-)^{\phi_{\vec{u}}(\vec{x})}, 
\end{equation}
where $\phi_{\vec{u}}(\vec{x})\equiv\phi(\vec{x})+\vec{u}\cdot\vec{x}$. 

Note that $\phi_{\vec{u}}(\vec{x})$---like the original $\phi(\vec{x})$---is \emph{also} a quadratic refinement of the bilinear form $(\vec{x}, \vec{y}) = \vec{x}^\intercal A \vec{y}$. This is because it also satisfies a version of equation~(\ref{cohomology}): 
\begin{equation}
    \phi_{\vec{u}}(\vec{x}+\vec{y})=\phi_{\vec{u}}(\vec{x})+\phi_{\vec{u}}(\vec{y})+\vec{x}^\intercal A \vec{y}
\label{cohomology2}
\end{equation}
The fact that equations~(\ref{cohomology}) and (\ref{cohomology2}) are unaffected by a linear shift in $\phi$ partly explains why the latter are called `quadratic refinements.'

We have found that introducing $\vec{u} \neq \vec{0}$ merely results in a linear shift in the quadratic refinement $\phi$. Therefore, computing $\braket{(-)^{\vec{u}}|G}$ follows the same steps as the computation of $\bra{+}^{\otimes M}\ket{G} = \braket{(-)^{\vec{0}}|G}$ in Section~\ref{sec: Arf_inv}, except for the replacement $\phi(\vec{x}) \to \phi_{\vec{u}}(\vec{x})$. 

Adapting \eqref{allfactors} to (\ref{amplitudephiu}), we find that $\braket{(-)^{\vec{u}}|G}$ equals:
\begin{equation}
    2^{-M}\!
    \prod_{k=1}^{\dim \ker A} \Bigg(1 + (-1)^{\phi_{\vec{u}}(\vec{w}_k)} \Bigg)
    \cdot\!\!\!\!
    \prod_{i=1}^{({\rm rank}A)/2}
    \Bigg(1 + (-1)^{\phi_{\vec{u}}(\vec{x}_i)} + (-1)^{\phi_{\vec{u}}(\vec{y}_i)} + (-1)^{1+\phi_{\vec{u}}(\vec{x}_i)+\phi_{\vec{u}}(\vec{y}_i)}\Bigg)
    \nonumber
\end{equation}
In the above expression, the symplectic basis vectors $\{\vec{w}_k\}_{k=1}^{\dim \ker A} \cup \{\vec{x}_i, \vec{y}_i\}_{i=1}^{({\rm rank} A)/2}$ are as defined below equation~(\ref{appeqn: A_new}) and in footnote~\ref{symplecticbasisfootnote} in the main text. Repeating the analysis of Section~\ref{sec: Arf_inv}, we find the following:
\begin{itemize}
\item $\braket{(-)^{\vec{u}}|G}\neq 0$ if and only if $\phi_{\vec{u}}(\vec{w}_k)=0$ for all $1 \leq k \leq \dim \ker A$. Said differently, the non-zero $X$-eigenbasis components of the wavefunction of $\ket{G}$ are in one-to-one correspondence with the solutions $\vec{u}$ of the linear equations: 
\begin{equation}
\vec{u}\cdot \vec{w}_k = \phi(\vec{w}_k) \qquad {\rm for}~1 \leq k \leq \dim \ker A
\label{nonvanishcond}
\end{equation}
There are $2^{({\rm rank} A)}$ such nonvanishing components.
\item If $\braket{(-)^{\vec{u}}|G}\neq 0$ then it equals $\pm 2^{-({\rm rank} A)/2}$.
\item Writing $\braket{(-)^{\vec{u}}|G}=(-)^{\Delta_{\vec{u}}}\times 2^{-({\rm rank} A)/2}$, the phase exponent $\Delta_{\vec{u}}$ is the Arf invariant of the quadratic refinement $\phi_{\vec{u}}(\vec{x})$: 
\begin{equation}
    \Delta_{\vec{u}}=\sum_{i=1}^{({\rm rank}A)/2} \phi_{\vec{u}}(\vec{x}_i)\,\phi_{\vec{u}}(\vec{y}_i)
\end{equation}
\end{itemize}

\section{Multi-invariants of stabilizer states}
\label{appendix:MI}
Multi-invariants were designed to characterize multi-partite entanglement structures in generic quantum states, initially for the purpose of exhibiting holographic duals of bulk geometric features in the AdS/CFT correspondence. Recent progress on multi-invariants can be found in \cite{Gadde_2024, Gadde2025, Gadde:2025ybn, gadde2025multipartiteentanglementmonotones} and in \cite{akella2026multiinvariantsstabilizerstates}, which is specifically devoted to multi-invariants of stabilizer states. 
 
This appendix studies multi-invariants of graph states. The analysis automatically extends to multi-invariants of all stabilizer states because every graph state is LC-equivalent to a graph state \cite{PhysRevA.69.022316} and multi-invariants are \emph{invariant} with respect to local unitary transformations. A partial motivation for studying multi-invariants of stabilizer states is to complement the results of \cite{akella2026multiinvariantsstabilizerstates} in the hope of understanding the information-theoretic utility of multi-invariants. We also wish to showcase and compare the two methods of computing partial amplitudes of graph states, which were developed in this paper: the generic method of Sections~\ref{sec: Innerprod}-\ref{sec: Arf_inv} and the more customized method of Section~\ref{sec:compute_permut}, which was designed specifically for multi-invariant calculations.

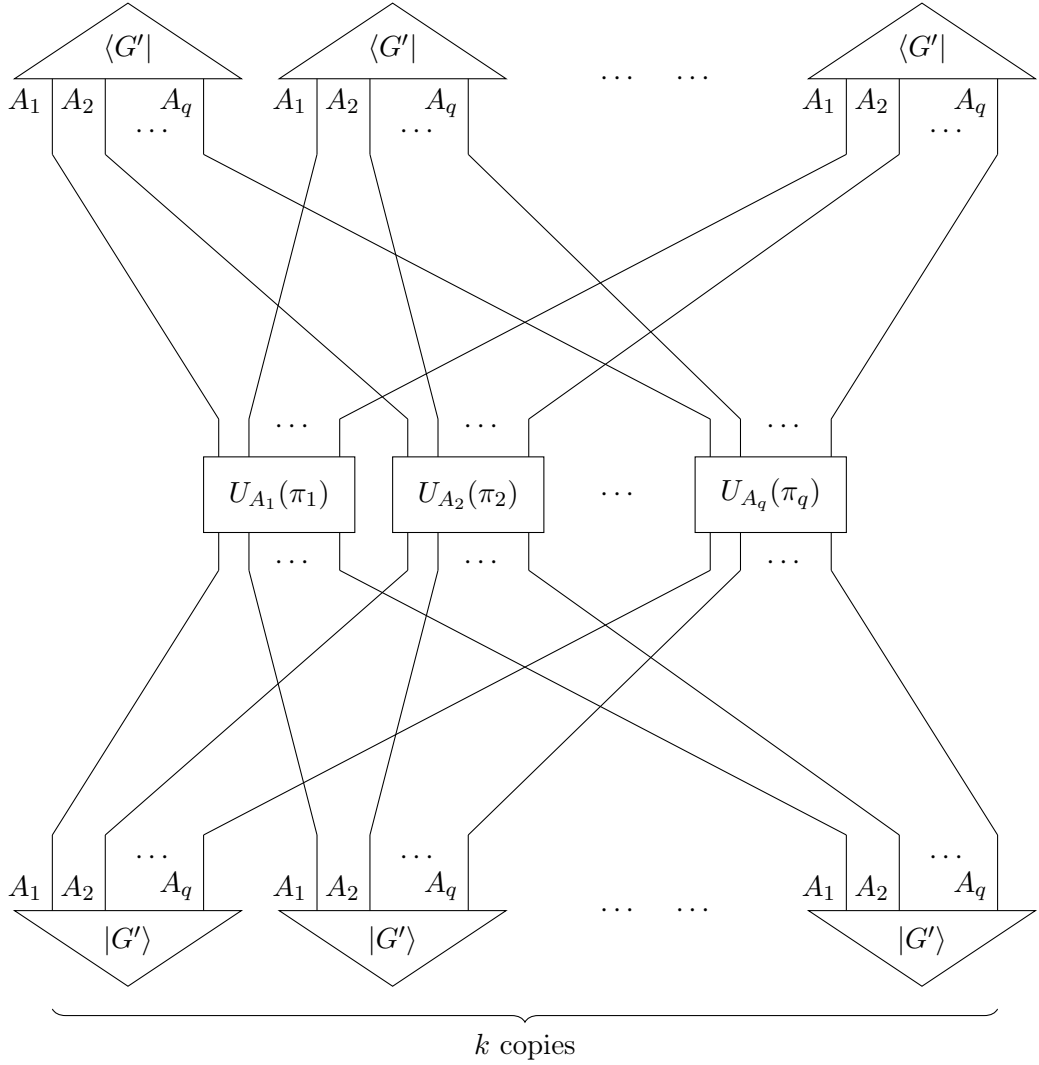
\begin{figure}[ht]
    \centering
    \begin{tikzpicture}[scale=1.0]
        \begin{scope}[shift={(0,0)}]
        \draw[thin] (0,-0.5) -- (-1.5,0.5) -- (1.5,0.5) -- cycle;
        \node at (0,0.1) {$\ket{G'}$};
        \draw[thin] (-1,0.5) -- (-1,1.5);
        \draw[thin] (-0.3,0.5) -- (-0.3,1.5);
        \draw[thin] (1,0.5) -- (1,1.5);
        \node at (0.35,1.2) {$\cdots$};
        \node[above left] at (-1,0.5) {$A_1$};
        \node[above left] at (-0.3,0.5) {$A_2$};
        \node[above left] at (1,0.5) {$A_q$};        

        \draw[thin] (3.5,-0.5) -- (2,0.5) -- (5.0,0.5) -- cycle;
        \node at (3.5,0.1) {$\ket{G'}$};
        \draw[thin] (2.5,0.5) -- (2.5,1.5);
        \draw[thin] (3.2,0.5) -- (3.2,1.5);
        \draw[thin] (4.5,0.5) -- (4.5,1.5);
        \node at (3.85,1.2) {$\cdots$};
        \node[above left] at (2.5,0.5) {$A_1$};
        \node[above left] at (3.2,0.5) {$A_2$};
        \node[above left] at (4.5,0.5) {$A_q$};

        \node at (6.5,0.5) {$\cdots$};
        \node at (7.5,0.5) {$\cdots$};
        \draw[thin] (10.5,-0.5) -- (9,0.5) -- (12.0,0.5) -- cycle;
        \node at (10.5,0.1) {$\ket{G'}$};
        \draw[thin] (9.5,0.5) -- (9.5,1.5);
        \draw[thin] (10.2,0.5) -- (10.2,1.5);
        \draw[thin] (11.5,0.5) -- (11.5,1.5);
        \node at (10.85,1.2) {$\cdots$};
        \node[above left] at (9.5,0.5) {$A_1$};
        \node[above left] at (10.2,0.5) {$A_2$};
        \node[above left] at (11.5,0.5) {$A_q$};

        \draw[thin] (-1,1.5) -- (1.2,5);
        \draw[thin] (-0.3,1.5) -- (3.7,5);
        \draw[thin] (1,1.5) -- (7.7,5);
        \draw[thin] (2.5,1.5) -- (1.6,5);
        \draw[thin] (3.2,1.5) -- (4.1,5);
        \draw[thin] (4.5,1.5) -- (8.1,5);
        \draw[thin] (9.5,1.5) -- (2.8,5);
        \draw[thin] (10.2,1.5) -- (5.3,5);
        \draw[thin] (11.5,1.5) -- (9.3,5);      
        \end{scope}

        \begin{scope}[shift={(0,12)}]
        \draw[thin] (0,0.5) -- (-1.5,-0.5) -- (1.5,-0.5) -- cycle;
        \node at (0,-0.1) {$\bra{G'}$};
        \draw[thin] (-1,-0.5) -- (-1,-1.5);
        \draw[thin] (-0.3,-0.5) -- (-0.3,-1.5);
        \draw[thin] (1,-0.5) -- (1,-1.5);
        \node at (0.35,-1.2) {$\cdots$};
        \node[below left] at (-1,-0.5) {$A_1$};
        \node[below left] at (-0.3,-0.5) {$A_2$};
        \node[below left] at (1,-0.5) {$A_q$};
        
        \draw[thin] (3.5,0.5) -- (2,-0.5) -- (5.0,-0.5) -- cycle;
        \node at (3.5,-0.1) {$\bra{G'}$};
        \draw[thin] (2.5,-0.5) -- (2.5,-1.5);
        \draw[thin] (3.2,-0.5) -- (3.2,-1.5);
        \draw[thin] (4.5,-0.5) -- (4.5,-1.5);
        \node at (3.85,-1.2) {$\cdots$};
        \node[below left] at (2.5,-0.5) {$A_1$};
        \node[below left] at (3.2,-0.5) {$A_2$};
        \node[below left] at (4.5,-0.5) {$A_q$};

        \node at (6.5,-0.5) {$\cdots$};
        \node at (7.5,-0.5) {$\cdots$};
        \draw[thin] (10.5,0.5) -- (9,-0.5) -- (12.0,-0.5) -- cycle;
        \node at (10.5,-0.1) {$\bra{G'}$};
        \draw[thin] (9.5,-0.5) -- (9.5,-1.5);
        \draw[thin] (10.2,-0.5) -- (10.2,-1.5);
        \draw[thin] (11.5,-0.5) -- (11.5,-1.5);
        \node at (10.85,-1.2) {$\cdots$};
        \node[below left] at (9.5,-0.5) {$A_1$};
        \node[below left] at (10.2,-0.5) {$A_2$};
        \node[below left] at (11.5,-0.5) {$A_q$};

        \draw[thin] (-1,-1.5) -- (1.2,-5);
        \draw[thin] (-0.3,-1.5) -- (3.7,-5);
        \draw[thin] (1,-1.5) -- (7.7,-5);
        \draw[thin] (2.5,-1.5) -- (1.6,-5);
        \draw[thin] (3.2,-1.5) -- (4.1,-5);
        \draw[thin] (4.5,-1.5) -- (8.1,-5);
        \draw[thin] (9.5,-1.5) -- (2.8,-5);
        \draw[thin] (10.2,-1.5) -- (5.3,-5);
        \draw[thin] (11.5,-1.5) -- (9.3,-5);     
        \end{scope}

        \begin{scope}[shift={(2,6)}]
            \draw[thin] (-1,0.5) -- (1,0.5) -- (1,-0.5) -- (-1,-0.5) -- cycle;
            \node at (0,0) {$U_{A_1}(\pi_1)$};
            \draw[thin] (-0.8,0.5) -- (-0.8,1);
            \draw[thin] (-0.4,0.5) -- (-0.4,1);
            \draw[thin] (0.8,0.5) -- (0.8,1);
            \node at (0.2,0.9) {$\cdots$};
            \draw[thin] (-0.8,-0.5) -- (-0.8,-1);
            \draw[thin] (-0.4,-0.5) -- (-0.4,-1);
            \draw[thin] (0.8,-0.5) -- (0.8,-1);
            \node at (0.2,-0.9) {$\cdots$};

            \draw[thin] (1.5,0.5) -- (3.5,0.5) -- (3.5,-0.5) -- (1.5,-0.5) -- cycle;
            \node at (2.5,0) {$U_{A_2}(\pi_2)$};
            \draw[thin] (1.7,0.5) -- (1.7,1);
            \draw[thin] (2.1,0.5) -- (2.1,1);
            \draw[thin] (3.3,0.5) -- (3.3,1);
            \node at (2.7,0.9) {$\cdots$};
            \draw[thin] (1.7,-0.5) -- (1.7,-1);
            \draw[thin] (2.1,-0.5) -- (2.1,-1);
            \draw[thin] (3.3,-0.5) -- (3.3,-1);
            \node at (2.7,-0.9) {$\cdots$};

            \node at (4.5,0) {$\cdots$};
            \draw[thin] (5.5,0.5) -- (7.5,0.5) -- (7.5,-0.5) -- (5.5,-0.5) -- cycle;
            \node at (6.5,0) {$U_{A_q}(\pi_q)$};
            \draw[thin] (5.7,0.5) -- (5.7,1);
            \draw[thin] (6.1,0.5) -- (6.1,1);
            \draw[thin] (7.3,0.5) -- (7.3,1);
            \node at (6.7,0.9) {$\cdots$};
            \draw[thin] (5.7,-0.5) -- (5.7,-1);
            \draw[thin] (6.1,-0.5) -- (6.1,-1);
            \draw[thin] (7.3,-0.5) -- (7.3,-1);
            \node at (6.7,-0.9) {$\cdots$}; 
        \end{scope}
        \draw[decorate,decoration={brace,amplitude=5pt,mirror}] (-1.0,-0.8) -- (11.5,-0.8);
        \node at (5.25,-1.3) {$k$ copies};
    \end{tikzpicture}
    \caption{A $k$-replica multi-invariant of a $q$-partite quantum state $\ket{G'}_{A_1 A_2 \dots A_q}$ involves a choice of $q$ permutations $\pi_i$, each acting on the $k$ replicas of party $A_i$.}
    \label{fig: defn_multiinv}
\end{figure}

In the most general terms, a $k$-replica multi-invariant of a $q$-partite quantum state $\ket{G'} \in \mathcal{H}_{A_1}\otimes\mathcal{H}_{A_2}\otimes \dots\otimes\mathcal{H}_{A_q}$ is (a function of) the following quantity:\footnote{The original definition of a multi-invariant~\cite{Gadde2025} assumes a particular choice of permutations $\pi_i$. We broaden their definition so as not to proliferate new jargon.}
\begin{equation}
\label{mugraphgeneral}
    \bra{G'}^{\otimes k}U_{A_1}(\pi_1)\otimes U_{A_2}(\pi_2)\otimes \dots\otimes U_{A_q}(\pi_q)\ket{G'}^{\otimes k}
    \equiv \bra{G'}^{\otimes k} U(\tilde{\pi}) \ket{G'}^{\otimes k}
\end{equation}
Here $U_{A_i}(\pi_i)$ is an appropriate representation of a permutation $\pi_i$, which permutes the $k$ replicas of the same party $A_i$. Note that---ignoring the conventional prefactor---the R{\'e}nyi entropy is a special case with $q=2$ and $\pi_1 = (1\, 2 \ldots k)$ and $\pi_2$ the identity. We illustrate quantity~\eqref{mugraphgeneral} in Figure~\ref{fig: defn_multiinv}. 

On the right hand side of equation~(\ref{mugraphgeneral}), we amalgamate the permutations $\pi_i$ into one big permutation $\tilde{\pi}$. This emphasizes that the computational technique developed in Section~\ref{sec:compute_permut} is precisely designed for quantities like~(\ref{mugraphgeneral}). On the other hand, we have seen in and above equation~(\ref{oldapproach}) that the more general approach of Sections~\ref{sec: Innerprod} and \ref{sec: Arf_inv} is also applicable. 

Specifically, assuming that $\ket{G'}$ is composed of $N$ qubits\footnote{Note that the $q$ parties which comprise $\ket{G'}$ may each contain multiple qubits.}, we define a new state $\ket{G}$ whose $\bra{++\ldots\,}$ partial amplitude equals the multi-invariant:
\begin{equation}
\label{eq: inner product of two large graph state}
\bra{G'}^{\otimes k} U(\tilde{\pi}) \ket{G'}^{\otimes k}
  = 
  \left(
  \bra{+}^{\otimes N}
    \prod_{i \sim j} CZ_{i,j}
  \right)^{\otimes k}
  U(\tilde{\pi})
  \left(
    \prod_{i \sim j} CZ_{i,j}
  \ket{+}^{\otimes N}
  \right)^{\otimes k} 
  \equiv \bra{+}^{\otimes N k}\ket{G}
\end{equation}
The state $\ket{G}$ is well defined because conjugating a $CZ$ gate by a permutation yields another $CZ$ gate and $\ket{+}^{\otimes Nk}$ is invariant under all permutations. Following the notation of the main text, we set $M = Nk$. 

\paragraph{Overview of multi-invariant calculations}
The tactic developed in Section~\ref{sec:compute_permut} tends to be more practical when the replica index $k$ and the permutations $\pi_i$ are meant to be varying and/or tunable. This is because in the approach of Sections~\ref{sec: Innerprod}-\ref{sec: Arf_inv} the $\pi_i$-dependence of the multi-invariant is buried in the implicit definition of the graph state $\ket{G}$; see equation~(\ref{eq: inner product of two large graph state}). We illustrate the usefulness of the `customized' computational technique in Appendix~\ref{sec: GHZ} by calculating the R{\'e}nyi multi-entropies \cite{Gadde_2022} and the dihedral invariants \cite{Harper:2025uui} of the three-qubit chain cluster state. 

On the other hand, when an explicit value of a particular multi-invariant is desired, the approach of Sections~\ref{sec: Innerprod}-\ref{sec: Arf_inv} can be more efficient. We illustrate this in Appendix~\ref{sec: C4} by computing the $n=2,3,4$ R{\'e}nyi multi-entropies of the four-qubit chain cluster state.

\subsection{Multi-invariants of the three-qubit chain cluster state}
\label{sec: GHZ}
We set $\ket{G'}$ to be the three-qubit chain cluster state. The latter is LC-equivalent to the three-party GHZ state, so calculations in this appendix also apply to $\ket{{\rm GHZ}_3}$. We follow the methodology developed in Section~\ref{sec:compute_permut}. 

\paragraph{R{\'e}nyi multi-entropy}
For later use in Appendix~\ref{sec: C4}, we start by defining the R{\'e}nyi multi-entropy for general $q$-partite states and only specialize to $q=3$ when it becomes necessary. The index-$n$ R{\'e}nyi multi-entropy is given by
\begin{equation}
    S_n^{(q)} = \frac{1}{(1-n)\, n^{q-2}} \log \mathcal{Z}_n^{(q)}\,,
\label{defmultirenyi}
\end{equation}
where 
\begin{equation}
    \mathcal{Z}_n^{(q)} 
    = \bra{G'}^{\otimes n^{q-1}}
    U_{A_1}(\pi_1)\otimes U_{A_2}(\pi_2)\otimes \ldots U_{A_{q-1}}(\pi_{q-1}) \otimes \mathbb{I}_{A_q} 
    \ket{G'}^{\otimes n^{q-1}}\,.
    \label{renyi-interm1}
\end{equation}
The permutations $\pi_i$ are chosen as follows. We envision the $n^{q-1}$ replicas of $\ket{G'}$ as living on a $(q-1)$-dimensional hypercube of size $n$, periodically identified. Figure~\ref{fig: nxnchessboard} displays the $q=3$ instance of this array of replicas. If we mark the replicas with $(q-1)$-tuples $\vec{h} \in \{0,\,1 \ldots n-1\}^{\otimes (q-1)}$ then the action of permutations $\pi_i$ is:
\begin{align}
\pi_1: & \quad (h_1,\, h_2 \ldots h_{q-1}) \to (h_1+1,\, h_2 \ldots h_{q-1}) \nonumber \\
\pi_2: & \quad (h_1,\, h_2 \ldots h_{q-1}) \to (h_1,\, h_2 + 1 \ldots h_{q-1}) \\
  & \qquad \qquad \cdots \nonumber \\
\pi_{q-1}: & \quad (h_1,\, h_2 \ldots h_{q-1}) \to (h_1,\, h_2 \ldots h_{q-1}+1) \nonumber
\end{align}
The additions are understood (mod~$n$). Note that the $q=2$ R{\'e}nyi multi-entropy is the ordinary R{\'e}nyi entropy.

It is convenient to start with equation~(\ref{eq:trpiG_result2}), which expresses~(\ref{renyi-interm1}) as a sum over stabilizers $O_{\vec{x}}$ of $\ket{G'}^{\otimes n^{q-1}}$. The delta functions in that sum demand that contributing stabilizers $O_{\vec{x}}$ act with an even number of Pauli $X$ and $Z$ operators in every cycle of the amalgamated permutation $\tilde\pi = \pi_1 \otimes \pi_2 \otimes \ldots \pi_{q-1} \otimes e$. Since all replicas of the $q^{\rm th}$ party make up their own individual cycles (this is the final $\ldots \otimes e$ in $\tilde\pi$), the contributing $O_{\vec{x}}$ must act as $I$ on each copy of $A_q$. 

Our discussion up to now applies to all $q$. We now specialize to $q=3$. 

\paragraph{Magic squares}
From equation~(\ref{graphstabgen}), the stabilizer group of the three-qubit chain cluster state is generated by $K_1 = X \otimes Z \otimes I$, $K_2 = Z \otimes X \otimes Z$ and $K_3 = I \otimes Z \otimes X$. Among the $2^3 = 8$ stabilizers of $\ket{G'}$, only the identity and $K_1$ act trivially on $A_3$. Therefore, the $O_{\vec{x}}$ that can contribute nontrivially to equation~(\ref{eq:trpiG_result2}) have a very simple structure: on each of the $n^{q-1} = n^2$ replicas, they act either with the identity or with $K_1$. Moreover, referring to equation~(\ref{tracesinglecycle}), we see that each contributing stabilizer adds $+1$ to the sum; the case with $-1$ never occurs because $X$ and $Z$ never show up in the same cycle of $\tilde\pi$.

\begin{figure}
    \centering
    \begin{tikzpicture}[scale=0.85]
        \draw[thin] (0,0) -- (10.3,0) -- (10.3,10) -- (0,10) -- (0,0);
        \draw[thin] (0,8) -- (10.3,8);
        \draw[thin] (0,6) -- (10.3,6);
        \draw[thin] (0,2) -- (10.3,2);
        \draw[thin] (2,0) -- (2,10);
        \draw[thin] (4,0) -- (4,10);
        \draw[thin] (8,0) -- (8,10);

        \node at (1,9) {$x_{(0,0)}$};
        \node at (3,9) {$x_{(0,1)}$};
        \node at (6,9) {$\cdots$};
        \node at (9.15,9) {$x_{(0,n-1)}$};
        \node at (1,7) {$x_{(1,0)}$};
        \node at (3,7) {$x_{(1,1)}$};
        \node at (6,7) {$\cdots$};
        \node at (9.15,7) {$x_{(1,n-1)}$};
        \node at (1,4) {$\vdots$};
        \node at (3,4) {$\vdots$};
        \node at (6,4) {$\ddots$};
        \node at (9.15,4) {$\vdots$};
        \node at (1,1) {$x_{(n-1,0)}$};
        \node at (3,1) {$x_{(n-1,1)}$};
        \node at (6,1) {$\cdots$};
        \node at (9.15,1) {$~x_{(n-1,n-1)}~$};
    \end{tikzpicture}
\caption{Stabilizers $O_{\vec{x}}$, which contribute to $\mathcal{Z}^{(3)}_{n}$ in summation~(\ref{eq:trpiG_result2}) are in one-to-one correspondence with magic squares---solutions of equations~(\ref{cond1}-\ref{cond2}).}
    \label{fig: nxnchessboard}
\end{figure}

In summary, the stabilizers $O_{\vec{x}}$ that contribute to equation~(\ref{eq:trpiG_result2}) =  (\ref{renyi-interm1}) either do or do not use $K_1$ on each site of the $n \times n$ grid shown in Figure~\ref{fig: nxnchessboard}. Let us denote the case where $K_1$ is (resp. is not) applied to the replica at $\vec{h} = (h_1, h_2) \equiv (i,j)$ with $x_{(i,j)} = 1$ (resp. $x_{(i,j)} = 0$). This assembles a binary string $\vec{x} \in \{0, 1\}^{\otimes n^2}$, which is laid out on the grid. Such length-$n^2$ bit strings are truncations of the summation index in equation~(\ref{eq:trpiG_result2}), which is a string of length $M = 3n^2$; the truncated bits stand for instances of $K_2$ and $K_3$, which are forbidden from contributing. To reiterate, potentially nontrivial summands in (\ref{eq:trpiG_result2}) can be represented by $n \times n$ arrays of 0's and 1's, as shown in Figure~\ref{fig: nxnchessboard}.

Eliminating $K_2$ and $K_3$ followed from imposing the selection criterion---each cycle must contain an even number of $X$- and $Z$-operators---on the $n^2$ cycles of $\pi_3 = e$. We must now impose the same condition on the $n$ cycles of $\pi_1$ and on the $n$ cycles of $\pi_2$. In the notation set up in the preceding paragraph, these conditions read:
\begin{align}
H_i & \equiv \sum_{j=0}^{n-1} x_{(i,j)} = 0 \qquad 0 \leq i \leq n-1 \label{cond1} \\
V_j & \equiv \sum_{i=0}^{n-1} x_{(i,j)} = 0 \qquad 0 \leq j \leq n-1 \label{cond2}
\end{align}
The sums are, of course, understood (mod~$2$). Generally, a grid of numbers whose rows and columns each add up to a specified sum is called a magic square. Evidently, to evaluate~(\ref{renyi-interm1}) is to count $n \times n$ magic squares over $\mathbb{F}_2$ with sum 0.

There is a single linear dependence among conditions~(\ref{cond1}) and (\ref{cond2}): $\sum_i H_i = \sum_j V_j$. Thus, the $n^2$ binary variables $x_{(i,j)}$ are subject to $2n -1$ linearly independent conditions. This leaves $n^2 - (2n-1) = (n-1)^2$ independent and unconstrained binary variables, which uniquely specify our magic squares. (Equivalently, \emph{every} $(n-1) \times (n-1)$ array of binary numbers can be uniquely completed to an $n \times n$ magic square.) Since all $2^{(n-1)^2}$ magic squares contribute $+1$ to (\ref{eq:trpiG_result2}), and since the number of cycles of $\tilde\pi$ is $r = n^2 + 2n$, we find
\begin{equation}
\mathcal{Z}_n^{(3)} =
\bra{G'}^{\otimes n^{2}} U_{A_1}(\pi_1)\otimes U_{A_2}(\pi_2) \otimes \mathbb{I}_{A_3} \ket{G'}^{\otimes n^{2}}
     = 2^{1-n^2}
\end{equation}
and, for the R{\'e}nyi multi-entropy:
\begin{equation}
S_n^{(3)} = \left(1+\frac{1}{n}\right) \log 2
\label{multirenyi3final}
\end{equation}

\begin{figure}
    \centering
    \begin{tikzpicture}[scale=0.8]
        \draw[thin] (0,0) circle (2.0);
        \draw[thin] (0,0) circle (3.0);
        \draw[thin] (0,0) circle (4.5);

        \draw[thin] (-60:2) -- (-60:4.5);
        \draw[thin] (-30:2) -- (-30:4.5);
        \draw[thin] (0:2) -- (0:4.5);
        \draw[thin] (30:2) -- (30:4.5);
        \draw[thin] (60:2) -- (60:4.5);
        \draw[thin] (90:2) -- (90:4.5);

        \node at (15:2.5) {$x_1$};
        \node at (15:3.75) {$x_{n+1}$};

        \node at (45:2.5) {$x_2$};
        \node at (45:3.75) {$x_{n+2}$};

        \node at (75:2.5) {$x_3$};
        \node at (75:3.75) {$x_{n+3}$};

        \node at (-15:2.5) {$x_n$};
        \node at (-15:3.75) {$x_{2n}$};

        \node at (-45:2.5) {$x_{n-1}$};
        \node at (-45:3.75) {$x_{2n-1}$};
        
        \node at (100:2.5) {$\cdot$};
        \node at (105:2.5) {$\cdot$};
        \node at (110:2.5) {$\cdot$};
        \node at (100:3.75) {$\cdot$};
        \node at (105:3.75) {$\cdot$};
        \node at (110:3.75) {$\cdot$};

        \node at (-70:2.5) {$\cdot$};
        \node at (-75:2.5) {$\cdot$};
        \node at (-80:2.5) {$\cdot$};
        \node at (-70:3.75) {$\cdot$};
        \node at (-75:3.75) {$\cdot$};
        \node at (-80:3.75) {$\cdot$};

        \node at (190:2.5) {$\cdot$};
        \node at (195:2.5) {$\cdot$};
        \node at (200:2.5) {$\cdot$};
        \node at (190:3.75) {$\cdot$};
        \node at (195:3.75) {$\cdot$};
        \node at (200:3.75) {$\cdot$};

    \end{tikzpicture}     
    \caption{Stabilizers $O_{\vec{x}}$, which contribute to $\mathcal{Z}_{2n}$ in summation~(\ref{eq:trpiG_result2}) are in one-to-one correspondence with ``magic annuli''---solutions of equations~(\ref{polar1}-\ref{radialcond}).}
    \label{fig: annulargrid}
\end{figure}
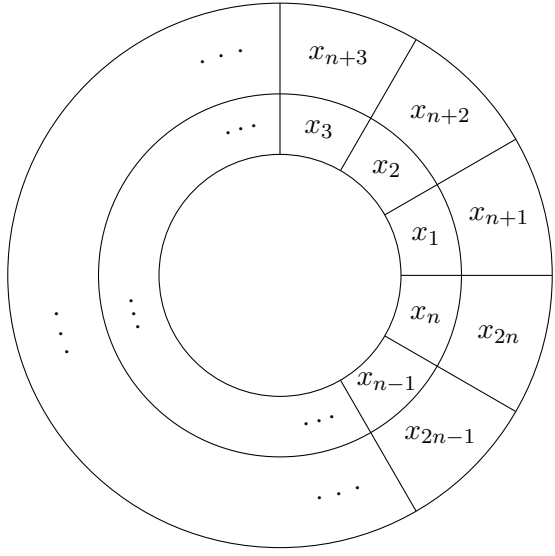

\paragraph{Dihedral invariants and magic annuli}
Reference~\cite{Gadde2025} defined the dihedral invariant as
\begin{equation} 
\label{eq:def_dihe_D}
    D_{2n} = \frac{1}{1-n} \log \frac{\mathcal{Z}_{2n}}{(\mathcal{Z}_2)^n}\,,
\end{equation}
where
\begin{equation}
\label{eq:def_dihe_Z}
    \mathcal{Z}_{2n} = 
    \bra{G'}^{\otimes 2n } U_{A_1}(\sigma_1)\otimes U_{A_2}(\sigma_2)\otimes \mathbb{I}_{A_3} \ket{G'}^{\otimes 2n} 
    \equiv \bra{G'}^{\otimes 2n } U(\tilde\sigma) \ket{G'}^{\otimes 2n}
\end{equation}
Similar to the previous discussion, we arrange the $2n$ replicas of $\ket{G'}$ on a $2 \times n$ annulus as shown in Figure~\ref{fig: annulargrid}. In the `coordinates' defined in the figure, the permutations $\sigma_1$ and $\sigma_2$ act as a rotation and a radial inversion, respectively:
\begin{align}
    \sigma_1 &\, = (1\, 2\, \dots\, n)\,(n\!+\!1\,\, n\!+\!2\, \ldots\, 2n) \nonumber \\
    \sigma_2 &\, = (1\,\,\, n\!+\!1)\,(2\,\,\,n\!+\!2)\dots (n\,\,\,2n)
\end{align}

We follow the same steps as in the R{\'e}nyi discussion and start with equation~(\ref{eq:trpiG_result2}). Once again, because parties $A_3$ are not permuted ($\sigma_3 = e$), every replica of $A_3$ makes up its own cycle of $\tilde\sigma$. Therefore, stabilizers $O_{\vec{x}}$ that contribute nontrivially in (\ref{eq:trpiG_result2}) must act as $I_{A_3}$ on each copy of $A_3$, which implies that they are assembled from $K_1$'s alone. Accordingly, we truncate from the summation index $\vec{x} \in \mathbb{F}_{2}^{6n}$ in (\ref{eq:trpiG_result2}) those $4n$ components, which are stand for applications of $K_2$ and $K_3$. With this truncation, stabilizers that may potentially contribute to (\ref{eq:trpiG_result2}) are represented by an annular array of 0's and 1's as shown in Figure~\ref{fig: annulargrid}.

We now impose the remaining conditions from the delta functions in (\ref{eq:trpiG_result2}): that every cycle of $\sigma_1$ and $\sigma_2$ contain an even number of Pauli $X$ and $Z$ operators. There are $n+2$ such conditions:
\begin{align}
        x_1+x_2+\dots + x_n\, &= 0 \label{polar1} \\
        x_{n+1}+x_{n+2}+\dots + x_{2n}\, &= 0 \label{polar2} \\
        x_i+x_{n+i}\, &=0 \qquad 1 \leq i \leq n \label{radialcond}
\end{align}
Naturally, the sums are understood (mod~$2$).

In an obvious extension of the nomenclature from Figure~\ref{fig: nxnchessboard}, solutions of (\ref{polar1}-\ref{radialcond}) define `magic annuli' over $\mathbb{F}_2$ with sum 0. There is one linear dependence among the conditions: the sum of all~(\ref{radialcond}) equals (\ref{polar1}) plus (\ref{polar2}). With $n+1$ linearly independent conditions on $2n$ entries in the array, we have exactly $2^{n-1}$ nonvanishing terms in (\ref{eq:trpiG_result2}). All of them contribute $+1$ to the sum because they arise from applications of $K_1 = X \otimes Z \otimes I$, which never place $X$ and $Z$ on the same qubit. All in all, we find:
\begin{align}
\bra{G'}^{\otimes 2n } U_{A_1}(\sigma_1)\otimes U_{A_2}(\sigma_2)\otimes \mathbb{I}_{A_3} \ket{G'}^{\otimes 2n} 
\,& = 2^{-2n+1} \\
D_{2n}\, &= \log 2
\label{dihedralfinal}
\end{align}
Results~(\ref{multirenyi3final}) and (\ref{dihedralfinal}) were previously derived in \cite{Gadde_2022,Harper:2025uui}.

\paragraph{Remarks}
We used equation~(\ref{eq:trpiG_result2}) and bypassed the derivations (\ref{eq:trpiG_resultFINAL}-\ref{eq: microscopic match}). Here engaging equation~(\ref{eq:trpiG_resultFINAL}) would be overkill because the permutations are really simple and all terms contribute positively to the summation. This is not so in general; see the final Remarks in Appendix~\ref{sec: C4}. One calculation, which showcases a full application of equation (\ref{eq:trpiG_resultFINAL}) is given in Section~\ref{sec:explicit}.

\subsection{R{\'e}nyi multi-entropies of the four-qubit chain cluster state}
\label{sec: C4}

In general, the technique of Sections~\ref{sec: Innerprod}-\ref{sec: Arf_inv} can be more efficient when one or a few concrete multi-invariants are sought. We illustrate this circumstance with an example, which involves the R{\'e}nyi multi-entropies of the four-qubit chain cluster state. The R{\'e}nyi multi-entropies for a general $q$-partite state are defined in the beginning of Appendix~\ref{sec: GHZ}. 

\begin{figure}[t]
    \centering
    \begin{subfigure}[b]{0.32\textwidth}
        \centering
        \includegraphics[width=\linewidth]{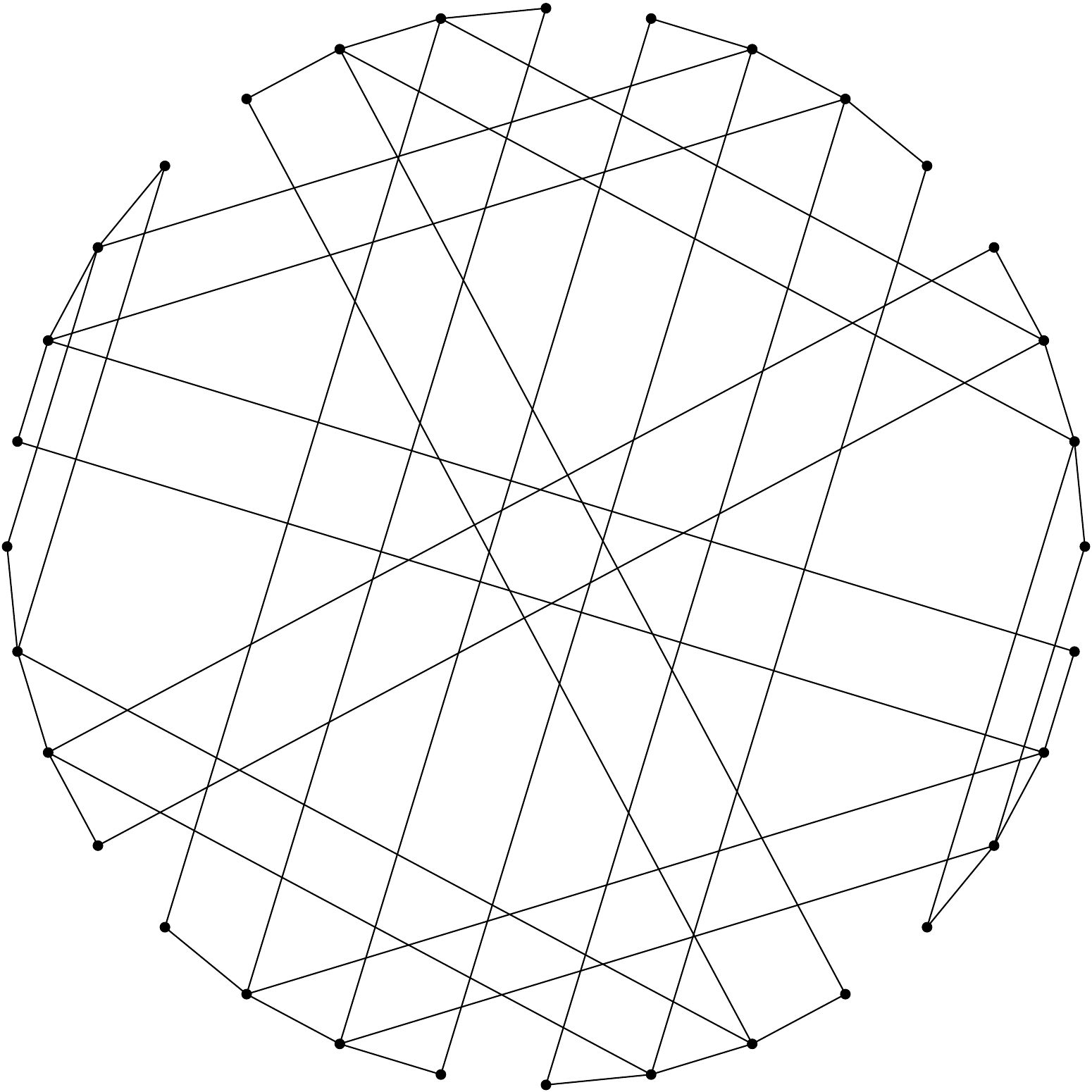}
        \caption{$n=2$: $\bra{+}^{32}\ket{G}=2^{-9}$}
        \label{fig:C4S24}
    \end{subfigure}
    \hfill
    \begin{subfigure}[b]{0.32\textwidth}
        \centering
        \includegraphics[width=\linewidth]{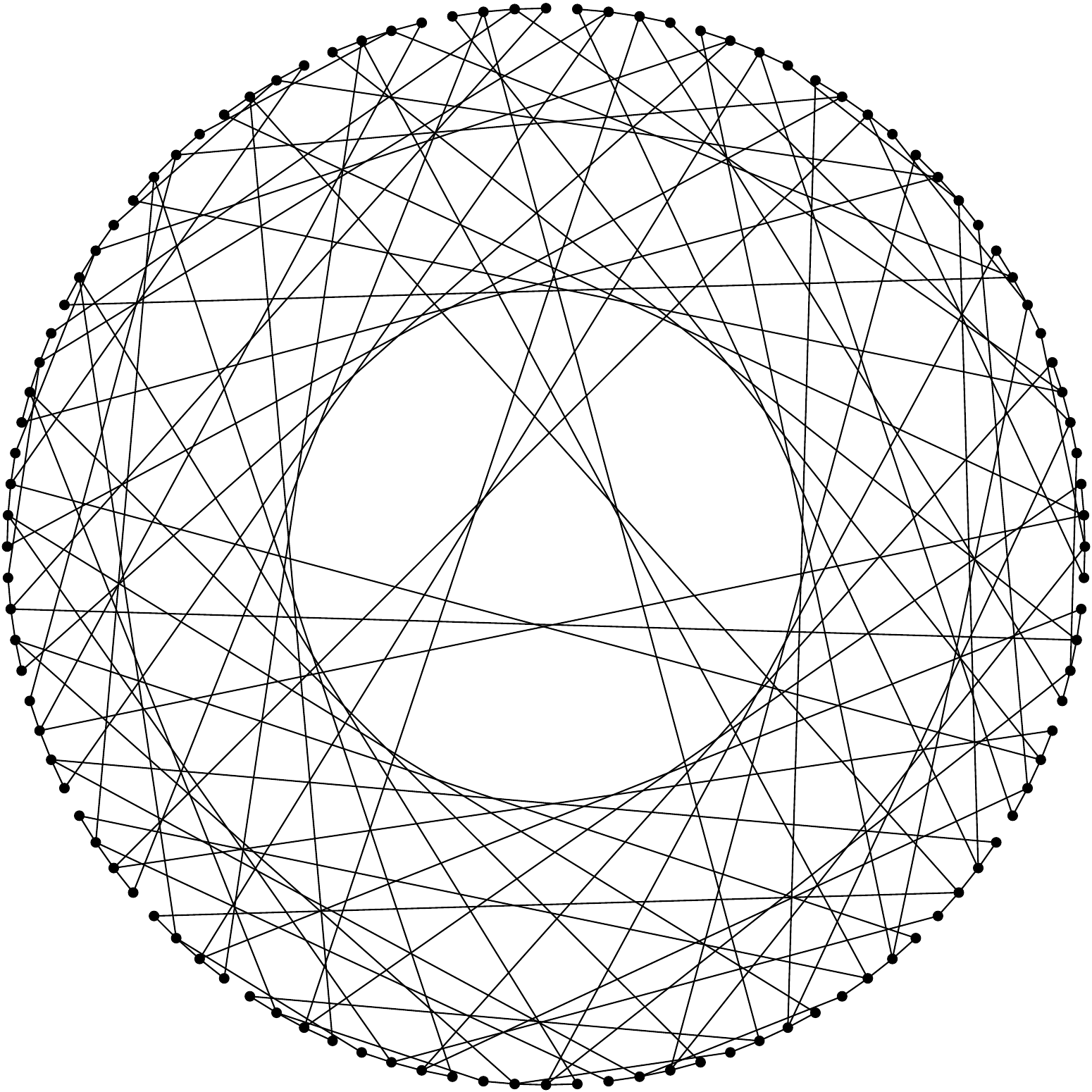}
        \caption{$n=3$: $\bra{+}^{108}\ket{G}=2^{-38}$}
        \label{fig:C4S34}
    \end{subfigure}
    \hfill
    \begin{subfigure}[b]{0.32\textwidth}
        \centering
        \includegraphics[width=\linewidth]{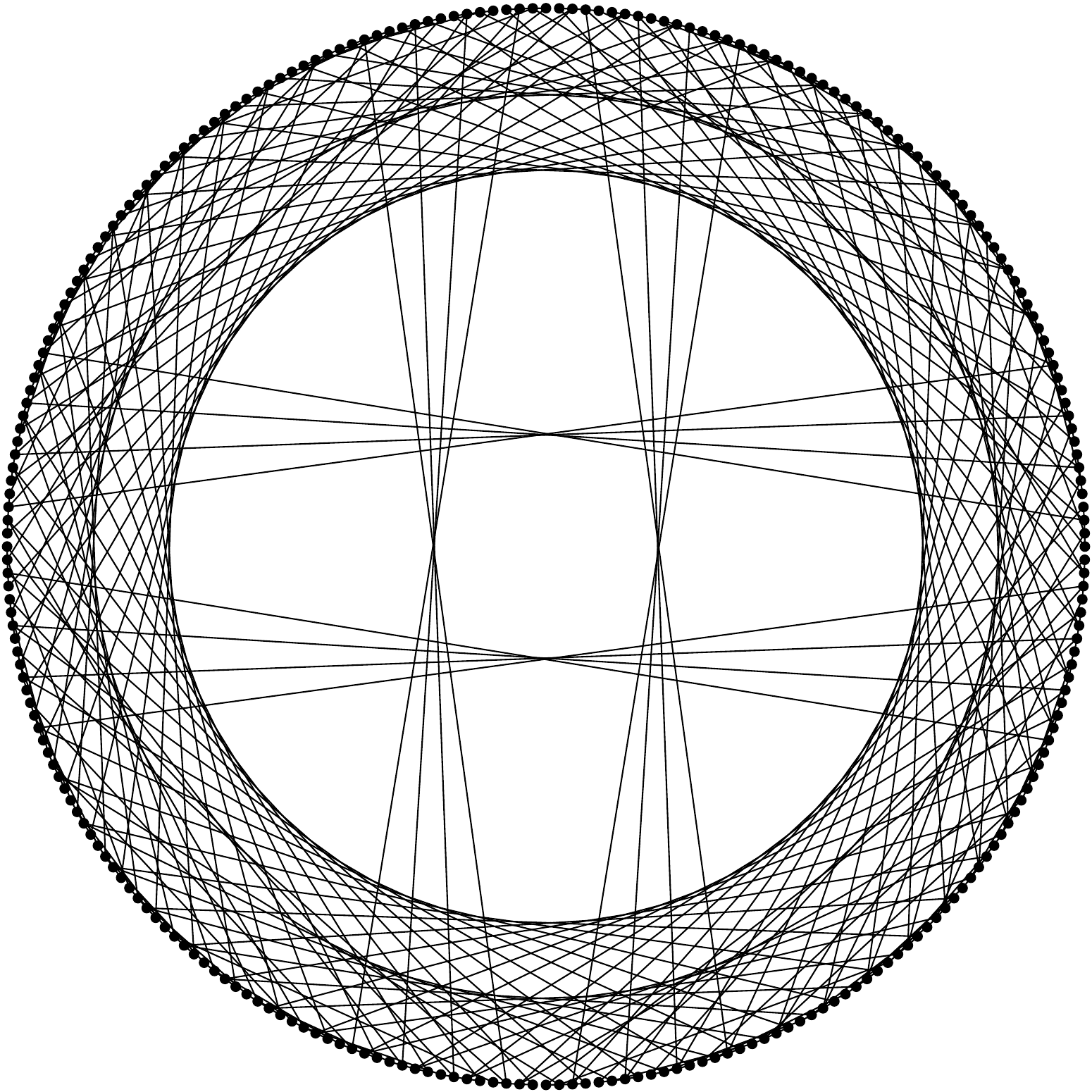}
        \caption{$n=4$: $\bra{+}^{256}\ket{G}=2^{-99}$}
        \label{fig:C4S44}
    \end{subfigure}

\caption{Graphs, which compute $\mathcal{Z}_n^{(4)}$ for $n=2,3,4$; see equation~(\ref{renyi4FINAL}).}    
    \label{fig:C4}
\end{figure} 

In still unpublished work \cite{treegraphME}, two of us and collaborators prove that the most fine-grained (one qubit per party) R{\'e}nyi multi-entropy of any $M$-qubit tree graph state is a polynomial in $(1/n)$ of degree $(M-2)$ or lower. Assuming this result, finding $S_n^{(4)}$ of the four-qubit chain cluster state requires knowing only three specific multi-entropies. For this purpose, the approach of Sections~\ref{sec: Innerprod}-\ref{sec: Arf_inv} is the most efficient.

\paragraph{Calculation}
Graphs $\ket{G}$ whose partial amplitude $\bra{+}^{\otimes 4n^3}\ket{G}$ computes $\mathcal{Z}_n^{(4)}$ for $n=2,3,4$ are shown in Figure~\ref{fig:C4}. The ranks of their adjacency matrices are 18, 76, and 198, respectively. In each case the Arf invariant is $\Delta = 0$.

Substituted into equation~(\ref{defmultirenyi}), these results translate into
\begin{equation}
S_{2}^{(4)}=(9/4) \log 2
\qquad {\rm and} \qquad
S_{3}^{(4)}=(19/9) \log 2
\qquad {\rm and} \qquad
S_{4}^{(4)}=(33/16) \log 2\,,
\label{renyis234}
\end{equation}
which are consistent with a unique quadratic in $1/n$:
\begin{equation}
S_{n}^{(4)}=\left(2+\frac{1}{n^2} \right) \log 2
\label{renyi4FINAL}
\end{equation}

\paragraph{Remarks} We used this calculation to illustrate the usefulness of the technique of Sections~\ref{sec: Innerprod}-\ref{sec: Arf_inv} for computing multi-invariants. Yet this demonstration should be taken with a grain of salt because we were only able to get to equation~(\ref{renyi4FINAL}) using the unpublished result from \cite{treegraphME}. In fact, that result is proven using the methodology from Section~\ref{sec:compute_permut}.

That said, it would not be practical to jettison the method of Sections~\ref{sec: Innerprod}-\ref{sec: Arf_inv} from multi-invariant calculations. As a test, we invite the reader to try and compute results~(\ref{renyis234}) from equation~(\ref{eq:trpiG_result2}) or (\ref{eq:trpiG_resultFINAL}). Unlike in Appendix~\ref{sec: GHZ}, this is an intricate combinatorial problem. The additional complication is that---after imposing the delta functions on the cycles of $\pi_4 \!=\! e$\,---the stabilizers $O_{\vec{x}}$ which contribute to (\ref{eq:trpiG_result2}) are generated by $K_1 = X \otimes Z \otimes I \otimes I$ and $K_2 = Z \otimes X \otimes Z \otimes I$. As the first and second party are being acted on by both $X$ and $Z$, the signs in equation~(\ref{eq:trpiG_result2}) are nontrivial and remarkably tricky to account for.

\bibliographystyle{JHEP}
\bibliography{biblio.bib}
\end{document}